\documentclass{article}

\PassOptionsToPackage{numbers,compress,sort}{natbib}
\usepackage[main, final]{neurips_2026}
\makeatletter
\renewcommand{\@noticestring}{Preprint of accepted paper, 40th Conference on Neural Information Processing Systems (NeurIPS 2026).}
\makeatother

\usepackage[utf8]{inputenc}
\usepackage[T1]{fontenc}
\usepackage{hyperref}
\usepackage{url}
\usepackage{xurl}
\usepackage{booktabs}
\usepackage{amsfonts}
\usepackage{nicefrac}
\usepackage{microtype}
\usepackage{xcolor}
\definecolor{linkdark}{rgb}{0.05,0.20,0.50}
\hypersetup{colorlinks=true,linkcolor=linkdark,citecolor=linkdark,urlcolor=linkdark,filecolor=linkdark}
\usepackage{graphicx}
\usepackage{amsmath,amssymb}
\usepackage{multirow}
\usepackage{array}
\usepackage{placeins}
\newcolumntype{L}[1]{>{\raggedright\arraybackslash}p{#1}}

\IfFileExists{generated/numbers.tex}{

}{}

\title{Asymmetric Phase Coding Audio Watermarking}
\author{%
  Guang Yang$^{1}$, Fengchen Liu$^{2,3,4}$, Amir Ghasemian$^{1}$,\\
  \textbf{Zhong Wang$^{2,5}$, Ninareh Mehrabi$^{6}$, Homa Hosseinmardi$^{1}$}\\
  $^{1}$University of California, Los Angeles,
  $^{2}$Lawrence Berkeley National Laboratory,\\
  $^{3}$University of California, Berkeley,
  $^{4}$Dakota State University,\\
  $^{5}$University of California, Merced,
  $^{6}$Meta%
}

\begin{document}

\maketitle

\begin{abstract}
The proliferation of deepfake audio challenges voice-based authentication, and passive forensic detectors are sensitive to evolving generative models and to real-world channel distortions. We propose Asymmetric Phase Coding (APC), a training-free cryptographic signing layer for audio that anyone holding the public key can verify offline, designed as a compact and auditable provenance primitive that can stand alone or be stacked with learned watermarks. APC combines Ed25519 digital signatures (EdDSA, FIPS 186-5; 64-byte signatures) with Reed--Solomon error correction, pseudo-random STFT phase-bin selection, and a redundant quantization-index-modulation (QIM) code on log-magnitude differences of adjacent bin pairs, yielding a compact, non-repudiable, blind-extractable watermark. A successful verification proves who signed the embedded message and that the message is intact. On 1{,}905 LibriSpeech \texttt{test-clean} clips (10\,s, 44.1\,kHz) under eight attack configurations (identity, 10\% and 20\% end-cropping, 8\,kHz low-pass, 16\,kHz round-trip resampling, FLAC, MP3 at 128\,kbps and OGG-Vorbis at 128\,kbps), APC verifies between $98.3\%$ and $99.8\%$ of clips on every condition at an identity PESQ of $3.82$ and accepts no wrong-key or unsigned clip. On 2{,}048 clips from 18 sets of 15 public corpora spanning music, meeting, telephone, multilingual and mostly vocoded speech, it verifies between $97.1\%$ and $99.4\%$. It also survives platform processing chains, repeated MP3 and OGG coding at 128\,kbps, cropping up to bounds predicted in closed form, integer time shifts through an optional decoder-side search, and the DNS64 and DeepFilterNet3 speech enhancers on most clips, and it verifies a signed 10\,s clip in 3.17\,ms on one CPU core. We compare APC with neural baselines (reproduced AudioSeal and WavMark, and published SilentCipher results), including a capacity-matched operating point, and show that it stacks with AudioSeal.
\end{abstract}

\section{Introduction}
\label{sec:intro}

Neural speech synthesis~\cite{huang2022neural} now reaches naturalness ratings close to those of recorded speech~\cite{shen2018natural}, and generated voices have been used in large-scale financial fraud~\cite{khanjani2022audio}. Passive forensic detectors~\cite{muller2022does, khanjani2022audio, yi2023audio, li2024antispoof} generalize poorly to attacks unseen in training and to in-the-wild audio, and channel and compression distortions degrade them further~\cite{asvspoof2021}. Proactive watermarking~\cite{bhowmik2001audio} is complementary: a verifiable token is embedded when genuine audio is captured or published, and its absence or failure flags the signal. A hiding channel alone, however, gives no non-repudiation, because anyone who knows the algorithm can produce a valid mark. Learned watermarks carry 16--40-bit symmetric messages, so attributing a clip to a signer needs a registry lookup, and a file-level signature such as a C2PA manifest~\cite{c2pa_spec} can be removed from the file, and its hash binding does not match a re-encoded copy.

\begin{figure}[t]
  \centering
  \includegraphics[width=0.97\linewidth]{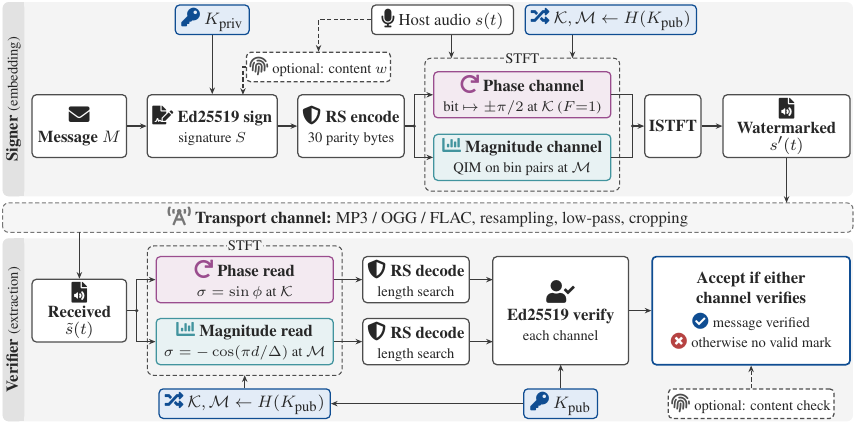}
  \caption{APC pipeline. The signer signs $M$ with $K_\text{priv}$ (Ed25519), encodes $\text{len}_{16}(M)\|M\|S$ with a Reed--Solomon code with 30 parity bytes, and writes it into a phase channel and a magnitude-QIM channel of one STFT. The verifier needs only $K_\text{pub}$ and accepts if either channel verifies. Dashed boxes show the optional content-bound mode: the signature also covers a feature $w$ of the host, and the verifier checks the feature of the received audio against it (Section~\ref{sec:security}). A verified message means that the signed message is intact, not that the audio is authentic (Section~\ref{sec:threat}).}
  \label{fig:pipeline}
\end{figure}

We propose Asymmetric Phase Coding (APC), which closes this gap with public-key cryptography (Fig.~\ref{fig:pipeline}). APC writes a complete Ed25519 signature into the waveform itself, so that anyone holding the public key can check offline who signed the embedded message, after the file has been compressed, resampled or cropped. APC signs a finished waveform and is indifferent to how it was produced: a microphone capture, an archival file and a generated clip are signed the same way. It is therefore not a model watermark but an infrastructure layer for AI provenance. Our contributions are:
\begin{enumerate}
\item A training-free signing layer with a specified format. An Ed25519 signature~\cite{rfc8032,nist2023digital} is protected by a Reed--Solomon code~\cite{reed1960,lin2004error} and written blindly into two parallel STFT channels: pseudo-random phase bins, and a quantization-index-modulation (QIM) code on log-magnitude differences of adjacent bins that survives lossy codecs. The format has a specification with conformance vectors and Python and Rust implementations (Section~\ref{sec:method}).
\item A precise guarantee. We separate forgery from erasure and state what a verification proves: the signer and the integrity of the message. An optional content-bound mode also binds the host audio and rejected every tested payload transplant, including one onto the nearest host in a 5{,}607-clip corpus (Section~\ref{sec:threat}).
\item A broad evaluation with a design analysis. On 1{,}905 LibriSpeech~\cite{panayotov2015librispeech} \texttt{test-clean} clips APC verifies $98.3$--$99.8\%$ under eight attacks, and on 2{,}048 clips from 18 corpus sets $97.1$--$99.4\%$, with 95\% intervals and wrong-key and unsigned controls that never accept. An ablation isolates what each design element contributes (Sections~\ref{sec:results}--\ref{sec:corpora}).
\item A mapped transport envelope. On 1{,}000 speech and 968 music clips we report where the mark survives: platform chains, repeated MP3 and OGG coding at 128\,kbps, loudness processing, packet loss and codecs from 96 or 128\,kbps. Closed-form bounds predict cropping and the minimum host duration, a decoder-side search recovers integer time shifts, and the mark survives the DNS64 and DeepFilterNet3 speech enhancers on most clips (Sections~\ref{sec:envelope} and~\ref{sec:neural}).
\end{enumerate}
APC needs no training corpus, no GPU and no generator-specific model. It suits C2PA-style pipelines~\cite{c2pa_spec,cai2024technical}, which sign at capture and bind content by hashes that re-encoding breaks; C2PA lists invisible watermarks as soft bindings~\cite{c2pa_spec}. The release contains the code, keys, clip lists and results behind every table and figure (Appendix~\ref{app:repro}). Code and data: \url{https://gy19a.github.io/asymmetric-phase-coding-audio-watermarking-js/}.

\section{Related work}
\label{sec:related}

\paragraph{Signal-processing and neural watermarks.} Classical audio watermarks trade capacity, imperceptibility and robustness~\cite{bhowmik2001audio}: LSB embedding breaks under requantization, echo hiding~\cite{gruhl1996echo} can add resonance, spread spectrum~\cite{cox1997} spreads the mark over many frequency bins, and phase coding~\cite{bender1996techniques,eurasip2015phase,ieee_adaptive_phase_coding,icassp2021phase} can be revealed by phase discontinuities~\cite{4460650}. Learned watermarks train encoder--decoder networks against channel attacks: WavMark~\cite{chen2023wavmark} reports PESQ $\approx4.21$, AudioSeal~\cite{sanroman2024audioseal} adds sample-level localization at PESQ $\approx4.47$, and SilentCipher~\cite{singh2024silentcipher} reports an SDR of about 47\,dB. They need training data and GPUs, carry 16--40-bit payloads rather than signatures, and face adversarial and distribution-shift concerns~\cite{adversarial_audio_survey}; timbre watermarking targets voice cloning~\cite{liu2024dear}. None provides cryptographic non-repudiation.

\paragraph{Provenance frameworks.} C2PA~\cite{c2pa_spec,c2pa_impl} specifies signed manifests and soft bindings, including invisible watermarks, without prescribing a watermarking algorithm, and CAI supports it~\cite{cai2024technical}. APC can be read as a signal-level carrier for a signed statement that survives the format changes in which file metadata, and with it an external manifest, is stripped.

\paragraph{Asymmetric watermarking.} Verifying a mark without handing over the means to forge it predates generative audio. Craver et al.~\cite{craver1998} showed that invisible marks alone cannot resolve ownership under counterfeit-original attacks, and Petitcolas et al.~\cite{petitcolas1998attacks} broke the marking systems of the day. Asymmetric schemes detect with public information and embed with private information~\cite{furon2000}; Eggers et al.~\cite{eggers2000} showed removal attacks against the public detectors then studied and left secure public detection open, and Cox, Do{\"e}rr and Furon distinguish cryptographic guarantees for embedded messages from resistance to removal~\cite{cox2006notcrypto}. The closest prior work is Steinebach and Dittmann~\cite{steinebach2003}, who combine digital signatures with fragile and invertible watermarking for publicly verifiable audio authentication. APC also carries a signature in the audio, but targets blind verification after lossy transport; its default verifier authenticates the signed message, and an optional mode also binds the content. We do not claim un-removability; we claim non-forgeability of the signed message (Section~\ref{sec:threat}), which a signature delivers directly, and measure the cost of erasure. The magnitude channel is a QIM code in the sense of Chen and Wornell~\cite{chen2001qim}.

\paragraph{Marks beyond the signal domain.} Recent work marks model parameters~\cite{ren2025p2mark}, neural codecs and their latents~\cite{zhou2025wmcodec,zhou2024traceablespeech}, the token stream of generative audio~\cite{milis2026hidden} and diffusion synthesis~\cite{liu2024groot}, or targets re-recording~\cite{liu2023dearRR}. None carries a public-key signature. Marks in the latent or token domain address neural-codec resynthesis, and Hidden in Plain Tokens retains detection after third-party codec re-encoding~\cite{milis2026hidden}; such marks are natural companions for a signal-domain signature. Whether independent marks can be stacked is itself a research question~\cite{petrov2025coexistence}, which we test in Section~\ref{sec:sota}. Appendix~\ref{app:related} gives a qualitative comparison table and a comparison with the first five of these works.

\section{Method}
\label{sec:method}

The inputs are a message $M$, an Ed25519 key pair $(K_\text{priv},K_\text{pub})$ and a host signal $s(t)$ (Fig.~\ref{fig:pipeline}). Appendix~\ref{app:format} lists every parameter.

\subsection{Payload}
\label{sec:crypto}
The signer signs the message and forms the payload from a 16-bit length prefix, the message and the 64-byte signature~\cite{rfc8032,bernstein2012ed25519}:
\begin{equation}
S=\text{Ed25519.Sign}(K_\text{priv},M),\qquad P=\text{len}_{16}(M)\,\|\,M\,\|\,S.
\end{equation}
$P$ is encoded with a Reed--Solomon code over $\mathrm{GF}(2^8)$ with 30 parity bytes, which corrects up to 15 byte errors~\cite{reed1960,lin2004error}. A 49-byte message gives $|P|{=}115$ bytes and a 145-byte (1{,}160-bit) codeword. The 16-bit length prefix is inside the code; the 32-bit per-channel stream header of Section~\ref{sec:extract} is not. Appendix~\ref{app:format} compares Ed25519 with RSA-1024.

\subsection{STFT, layout and phase channel}
\label{sec:embed}
The host is transformed by a rectangular-window, non-overlapping STFT with $N_\text{FFT}{=}2048$,
\begin{equation}
X(i,k)=A(i,k)\,e^{j\phi(i,k)},
\end{equation}
and frames are grouped by $G{=}8$ (0.37\,s at 44.1\,kHz). A public layout seed $\kappa=\mathrm{int}(\text{SHA-256}(K_\text{pub})[0{:}8])\bmod 2^{32}$ selects everything. A generator seeded with $\kappa$ shuffles the phase bins $\mathcal{K}\subset[60,300)$ (1.29--6.46\,kHz), a mid band chosen to survive resampling and low-pass filtering; lossy codecs damage it. Because the layout is public, extraction is blind. Each coded bit $d_n$ sets a phase target, and the offset to it at the anchor frame $i_0$ of the group is added at bin $k$:
\begin{equation}
\phi_\text{data}[n]=\begin{cases}+\pi/2, & d_n=1,\\ -\pi/2, & d_n=0,\end{cases}\qquad \Delta\phi[n]=\phi_\text{data}[n]-\phi(i_0,k).
\end{equation}
The signal is rebuilt as $s'(t)=\text{ISTFT}(A\,e^{j\phi'})$, where $\phi'=\phi+\Delta\phi$ at the written cells and $\phi'=\phi$ elsewhere. The offset is written into the anchor frame only ($F{=}1$): the decoder reads one frame per group, so writing all $G$ frames would add distortion without adding evidence (Section~\ref{sec:ablation}).

\subsection{Magnitude-QIM channel}
\label{sec:mag}
Lossy codecs redistribute phase inside critical bands while roughly keeping magnitudes, so a second channel carries the same payload in magnitudes. A permutation seeded with $(\kappa\oplus\texttt{0xDEADBEEF})\bmod2^{32}$ orders the fixed adjacent bin pairs $(k,k{+}1)$ of the band $[100,340)$ (2.15--7.32\,kHz). For a pair, the group-averaged log magnitudes and their difference are
\begin{equation}
\ell_j=\frac1G\sum_{i\in g}\ln\big(A(i,k_j)+\varepsilon\big),\quad j\in\{1,2\},\qquad d=\ell_1-\ell_2,
\end{equation}
with $\varepsilon{=}10^{-8}$. A bit $b$ snaps $d$ to the nearest cell of a quantizer with step $\Delta{=}1.0$\,nat whose parity is $b$:
\begin{equation}
c=\lfloor d/\Delta\rceil,\qquad c'=\begin{cases}c, & c\bmod 2=b,\\ c+1, & c\bmod 2\neq b \text{ and } d\ge c\Delta,\\ c-1, & \text{otherwise}.\end{cases}
\end{equation}
The shift $A(i,k_{1,2})\leftarrow A(i,k_{1,2})e^{\pm(c'\Delta-d)/2}$ realizes $d^*=c'\Delta$ and preserves $\ell_1+\ell_2$, both up to the floor $\varepsilon$. The decoder uses the soft values
\begin{equation}
\sigma_n=-\cos(\pi d/\Delta)\ \text{(magnitude)},\qquad \sigma_n=\sin\phi(i_0,k)\ \text{(phase)}.
\end{equation}

\paragraph{Replication and layout.} Each channel writes the 32-bit header three times and repeats the body
\begin{equation}
r=\max\big(1,\min(R,\lfloor(C-96)/|b|\rfloor)\big)
\end{equation}
times, where $C$ is the channel capacity, $|b|$ the body length and $R$ an upper bound ($R_p{=}1$, $R_m{=}5$). Soft values are summed over copies before one threshold, then RS-decoded, so redundancy works in APC's favor: the two magnitude copies add their evidence, and averaging each bit over the 8 frames of its group lets an exact one-frame shift still verify (Section~\ref{sec:envelope}). A 10\,s clip has 26 groups: the phase copy occupies 6 of them (the first 2.23\,s) and the two magnitude copies 21 (the first 7.80\,s); later groups carry no payload bits (Fig.~\ref{fig:layout} in Appendix~\ref{app:format}).

\subsection{Blind verification and minimum duration}
\label{sec:extract}
The verifier regenerates the layout from the public key, reads both channels, and accepts if either yields an RS codeword whose Ed25519 signature verifies; everything else means ``no valid watermark''. The stream header is outside the RS parity, so the verifier does not depend on it: it first tries the header value and otherwise searches all 160 admissible message lengths (0--159 bytes) on each channel, accepting only a payload whose RS decoding succeeds and whose signature verifies. The minimum host length is closed-form: both channels must carry $96{+}1{,}160=1{,}256$ bits; the magnitude channel carries 120 bits per group and binds at 11 groups, or 180{,}224 samples (4.09\,s). A sample-exact sweep confirms the floor: at 180{,}223 samples the embedder refuses all 30 test clips, and at 180{,}224 samples all 30 are signed and verify (Fig.~\ref{fig:duration} in Appendix~\ref{app:envelope}).

\subsection{Design elements}
\label{sec:v1}
Four design elements follow from reading the format against its decoder. The phase offset is written into one frame per group, the frame the decoder reads. The verifier does not depend on the stream header, which lies outside the RS parity; a verifier that stops at an out-of-range length can be gated by changing a few header bins (Section~\ref{sec:ablation}). The last partial frame of each clip (680 samples, 15.42\,ms, for 10\,s) is copied through rather than zeroed. Bits whose magnitudes fall below $\tau{=}10^{-6}$ are erased rather than decoded as data. Each element can be switched off, and Section~\ref{sec:ablation} measures it against the original default.

\section{Threat model and guarantee}
\label{sec:threat}

The layout is public, so an attacker can locate every watermarked bin; APC targets authenticity, not confidentiality.
Forgery is producing a valid signature for a message not previously signed under the legitimate key. Ed25519 is EUF-CMA secure under the elliptic-curve discrete-logarithm assumption in the random oracle model~\cite{brendel2021ed25519}, and knowing the layout does not bypass signature verification~\cite{rfc8032,bernstein2012ed25519}.
Erasure is producing a signal that fails verification. It is always possible at some cost in quality, so APC does not claim un-removability (Section~\ref{sec:limits}).

\paragraph{What a verification proves.} A successful verification proves that the embedded message was signed by the holder of $K_\text{priv}$ and has not been altered. The signature covers the message, not the waveform, so the default verifier authenticates the signer and the message rather than the surrounding audio. An optional content-bound mode also signs a secure sketch of a low-band content feature of the host, so that a payload moved into other audio no longer verifies (Section~\ref{sec:security}, Appendix~\ref{app:binding}). The main results use the default verifier.

\paragraph{Scope.} We separate resilience to innocuous transport (codecs, resampling, filtering, cropping) from robustness to a motivated attacker; Sections~\ref{sec:results}--\ref{sec:neural} measure the first and Section~\ref{sec:security} the second. Appendix~\ref{app:corr} maps our transformations to those of AudioSeal~\cite{sanroman2024audioseal}, AudioMarkBench~\cite{adversarial_audio_survey}, DeAR~\cite{liu2023dearRR} and ASVspoof 2019 PA~\cite{asvspoof2019}. The analog hole is outside the model: audio played through a loudspeaker and recaptured is a new recording, and APC authenticates a signed bitstream, not the semantic content of a sound. Unlike file metadata, which re-encoders strip silently, the mark survives the transport steps of Sections~\ref{sec:results}--\ref{sec:envelope}, so its absence after such steps is informative.

\section{Experiments}
\label{sec:exp}

\subsection{Setup}
\label{sec:dataset}
\label{sec:protocol}
\label{sec:attacks}
The main benchmark is built from LibriSpeech \texttt{test-clean}~\cite{panayotov2015librispeech} (CC BY 4.0; 40 speakers, 87 chapters, 2{,}620 utterances). Per speaker and chapter we concatenate utterances in id order, resample from 16 to 44.1\,kHz (soxr, high quality), quantize to 16-bit PCM and cut consecutive non-overlapping 10\,s windows, discarding each chapter's final partial window. This deterministic build yields 1{,}905 clips, all evaluated; the clip list is released (Appendix~\ref{app:data}). One fixed key pair signs every clip, a second serves as the wrong key, and a fixed 49-byte ASCII message is embedded (Appendix~\ref{app:protocol}). Signed audio is quantized to PCM16 before each attack. The eight benchmark attacks are identity, MP3 and OGG-Vorbis at 128\,kbps, FLAC (FFmpeg 7.0.2), 44.1$\rightarrow$16$\rightarrow$44.1\,kHz resampling, an 8\,kHz zero-phase Butterworth low-pass, and 10\% and 20\% end-cropping with the tail zeroed. Quality metrics compare the attacked watermarked clip with the original, so they include the embedding distortion; PESQ is wideband at 16\,kHz on the whole clip. Verification rates carry 95\% Wilson or exact Clopper--Pearson intervals; paired differences carry paired bootstrap 95\% intervals (2{,}000 resamples) and exact McNemar tests.

\subsection{Main benchmark}
\label{sec:results}

\begin{table}[t]
\centering
\caption{Main benchmark: APC on 1{,}905 LibriSpeech \texttt{test-clean} clips (10\,s, 44.1\,kHz). Entries are verified clips and verification rates with 95\% Wilson intervals for PCM16 output, the rate for unquantized output, and the mean PESQ-WB of the attacked clip against the original. No wrong-key (30{,}480) or unsigned (3{,}810) verification was accepted.}
\label{tab:results}
\footnotesize
\setlength{\tabcolsep}{4pt}
\IfFileExists{generated/tab_main.tex}{\begin{tabular}{@{}lcccc@{}}
\toprule
Attack & Verified (of 1{,}905) & Rate [95\% CI] & Unquantized & PESQ \\
\midrule
Identity & 1{,}902 & 0.998 [0.995, 0.999] & 1.000 & 3.82 \\
MP3, 128\,kbps & 1{,}900 & 0.997 [0.994, 0.999] & 0.997 & 3.82 \\
OGG Vorbis, 128\,kbps & 1{,}873 & 0.983 [0.976, 0.988] & 0.994 & 3.82 \\
FLAC & 1{,}902 & 0.998 [0.995, 0.999] & 0.999 & 3.82 \\
Resampling via 16\,kHz & 1{,}902 & 0.998 [0.995, 0.999] & 0.999 & 3.82 \\
End crop 10\% & 1{,}902 & 0.998 [0.995, 0.999] & 1.000 & 2.58 \\
End crop 20\% & 1{,}902 & 0.998 [0.995, 0.999] & 1.000 & 2.04 \\
Low-pass 8\,kHz & 1{,}902 & 0.998 [0.995, 0.999] & 1.000 & 3.82 \\
\midrule
\multicolumn{5}{@{}l}{Identity quality: PESQ 3.82, SNR 21.4\,dB, STOI 0.996, LSD 8.6\,dB} \\
\bottomrule
\end{tabular}
}{\textbf{[generated/tab\_main.tex missing]}}
\end{table}

\begin{figure}[t]
  \centering
  \includegraphics[width=\linewidth]{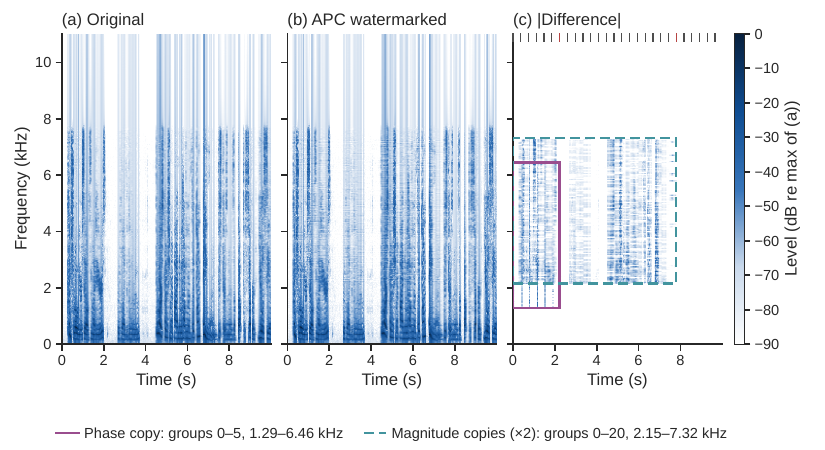}
  \caption{Spectrograms of a 10\,s LibriSpeech \texttt{test-clean} clip (a) before and (b) after signing with APC (wideband profile, 49-byte message), and (c) the magnitude of their difference. All panels use the coder's own analysis (2048-sample rectangular frames, no overlap) on one dB scale relative to the largest bin of (a). Tick marks on the top edge of (c) are group boundaries (8 frames, 0.37\,s). The phase copy occupies groups 0--5 (first 2.23\,s) and writes only the first frame of each group; the two magnitude copies occupy groups 0--20 (first 7.80\,s). Frames after the last used group are unchanged.}
  \label{fig:spectrogram}
\end{figure}

APC verifies between $0.983$ (OGG) and $0.998$ of the 1{,}905 clips (Table~\ref{tab:results}); without PCM16 quantization it verifies every clip under identity, low-pass and both crops. Identity PESQ is $3.82$ and SNR $21.4$\,dB, and the embedding changes only the groups that carry payload (Fig.~\ref{fig:spectrogram}); cropping lowers PESQ because the removed tail counts as distortion against the original. The two channels complement each other: the phase channel carries most verifications, and the magnitude channel verifies 53 MP3 and 189 OGG clips on which the phase channel fails.

\subsection{What each design element contributes}
\label{sec:ablation}
\label{sec:fixes}

\begin{table}[t]
\centering
\caption{Ablation on the 1{,}905 clips of Table~\ref{tab:results}. Each row switches one design element back to its original default; entries are APC minus the ablated coder: identity PESQ with a paired bootstrap 95\% interval, the range of verification differences over the eight attacks, and selected counts of discordant clips (APC only / ablated only). No arm accepted a wrong-key or unsigned clip.}
\label{tab:ablation}
\footnotesize
\setlength{\tabcolsep}{4pt}
\IfFileExists{generated/tab_ablation.tex}{\begin{tabular}{@{}llccc@{}}
\toprule
Design element & Original & $\Delta$PESQ [95\% CI] & $\Delta$verify (pp) & Discordant \\
\midrule
Single-frame write & 8-frame write & \ensuremath{+}0.432 [0.427, 0.437] & \ensuremath{-}0.4 to \ensuremath{+}0.2 & MP3 4/1, OGG 2/9 \\
Tail passthrough & tail zeroed & \ensuremath{+}0.201 [0.197, 0.205] & 0 (all 8) & 0/0 \\
Header-independent search & header-trusting & 0 (verifier only) & \ensuremath{+}0.7 to \ensuremath{+}0.8 & 13--16/0 \\
Silent-bin erasure & decoded & 0 (verifier only) & \ensuremath{+}0.3 to \ensuremath{+}0.4 & 5--7/0 \\
\bottomrule
\end{tabular}
}{\textbf{[generated/tab\_ablation.tex missing]}}
\end{table}

Table~\ref{tab:ablation} switches one element off at a time. The single-frame phase write accounts for most of the quality ($+0.432$ PESQ, $+7.2$\,dB SNR) and tail passthrough for another $+0.201$; against the original defaults as a whole, APC raises identity PESQ by $0.600$ [$0.594$, $0.606$]. The verifier-side elements leave the signed audio unchanged and add verifications: header-independent search $+0.7$ to $+0.8$ points and silent-bin erasure $+0.3$ to $+0.4$ points over the eight attacks. Header independence also matters for security. A verifier that trusts the stream header can be gated: changing three phase bins and three magnitude groups of the header drives such a verifier from $1.000$ to $0.000$ on 30 TED-LIUM clips with a PESQ change below $10^{-4}$. APC's verifier closes this gate by construction, and a wrong candidate length is accepted only with a valid signature (none of 30{,}480 wrong-key and 3{,}810 unsigned verifications). The header also governs head crops: the phase copy sits in the first 6 of 26 groups (2.23\,s), and APC verifies $0.992$, $0.983$, $0.970$, $0.963$ and $0.923$ after the first 5, 10, 20, 30 and 40\% of the clip are zeroed, where a header-trusting verifier gives $0$ from 5\% on (Appendix~\ref{app:quality}). The QIM step offers a further quality lever: at $\Delta{=}0.5$ APC reaches PESQ $4.11$ (Appendix~\ref{app:oppoint}).

\subsection{Generality across corpora and music}
\label{sec:corpora}

We ran APC on 2{,}048 clips from 18 sets of 15 public corpora (Appendix~\ref{app:data}): 968 music clips~\cite{defferrard2017fma,tzanetakis2002gtzan,rafii2019musdb18hq}, 180 clips of clean read and lecture speech~\cite{panayotov2015librispeech,hernandez2018tedlium3} and 900 clips from eleven further sets~\cite{carletta2006ami,gemmeke2017audioset,ardila2019commonvoice,thiemann2013demand,delrio2022earnings,conneau2023fleurs,galvez2021peoples,simsamu2025,chung2020voxconverse,frank2021wavefake}. APC verifies between $0.971$ (OGG) and $0.994$ (identity), and no wrong-key or unsigned clip was accepted (Fig.~\ref{fig:corpora} in Appendix~\ref{app:corpora}). Identity verification is $0.998$ on music, $1.000$ on clean speech and $0.988$ on the further corpora; six of those eleven verify every clip, including both AMI recordings of overlapping meetings on headset and distant microphones. Narrowband capture has its own profile: on 72 simulated emergency-medical calls at 8\,kHz (SIMSAMU), the narrowband profile (magnitude band 345--3618\,Hz, selected by the signer; the verifier tries both profiles) verifies all 72 clips under all eight attacks, at an identity PESQ cost of $0.25$. On the 968 music clips, the structural envelope of Section~\ref{sec:envelope} behaves as on speech: the tail crop bound is the same, head crops of 5--40\% verify $0.978$--$0.993$, and the decoder-side search recovers every tested shift ($0.991$--$0.998$). Appendices~\ref{app:corpora} and~\ref{app:envelope} give every corpus and condition.

\subsection{Transport envelope}
\label{sec:envelope}

\begin{figure}[tp]
  \centering
  \includegraphics[width=\linewidth]{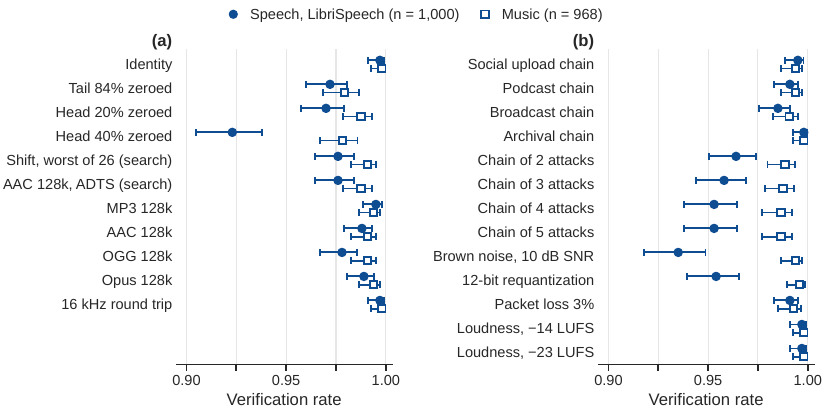}
  \caption{APC verification under transport conditions beyond the eight benchmark attacks, on 1{,}000 LibriSpeech \texttt{test-clean} speech clips and 968 music clips (PCM16 after signing), with 95\% Wilson intervals. Table~\ref{tab:envelope} in Appendix~\ref{app:envelope} lists every condition; no wrong-key or unsigned verification was accepted.}
  \label{fig:envelope}
\end{figure}

\begin{figure}[t]
  \centering
  \includegraphics[width=\linewidth]{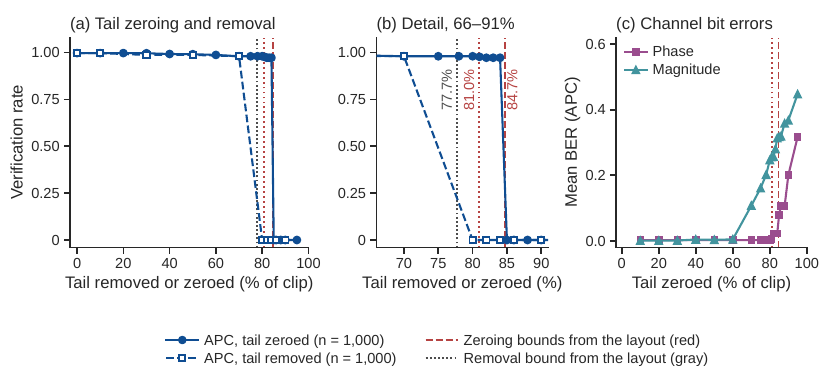}
  \caption{Tail cropping on 1{,}000 LibriSpeech \texttt{test-clean} clips (10\,s, PCM16). Circles: the last $p$\% of the clip is set to zero; squares: it is removed. Error bars are 95\% Wilson intervals. The red lines are derived from the layout, not fitted: the last phase group is read in frame 40, which ends before sample 83{,}968, so zeroing up to 81.0\% keeps every phase bit, and RS parity absorbs the loss of that group (7 bytes), extending the bound to 84.7\%. Removal is bounded by the length of one phase copy (98{,}304 samples, gray line).}
  \label{fig:bounds}
\end{figure}

Fig.~\ref{fig:envelope} and Table~\ref{tab:envelope} map the transport envelope on 1{,}000 benchmark clips and on the 968 music clips (Appendix~\ref{app:envelope}). Cropping is predicted by the layout (Fig.~\ref{fig:bounds}): a tail crop of up to 81.0\% keeps every phase bit, and RS parity extends the bound to 84.7\%; APC verifies $0.972$ on speech and $0.979$ on music with 84\% of the clip zeroed, just inside the predicted edge. Translations are recoverable. Without a search, a delay of one sample and of exactly one frame verify, because the magnitude reader averages each bit over the 8 frames of its group; an optional decoder-side search over sample offsets and frame skips recovers all 26 tested shifts ($0.976$--$0.997$) and a 44-sample advance ($0.859$) without changing the stream. The same search recovers raw ADTS AAC at 128\,kbps ($0.976$), whose encoder delay an MP4 container records and a raw stream does not. Codecs and processing chains keep the mark: MP3 and Opus from 96\,kbps and AAC and OGG from 128\,kbps verify $0.963$--$0.997$, upload, podcast, broadcast and archive chains $0.985$--$0.998$, chains of two to five benchmark attacks $0.953$--$0.964$, and five generations of MP3 at 128\,kbps $0.979$ on speech and $0.975$ on music. On speech, loudness normalization, compression, equalization and gain changes keep APC at $0.968$ or more, 3\% packet loss at $0.991$, 12-bit requantization at $0.954$, and brown noise at 10\,dB SNR at $0.935$.

\subsection{Neural speech enhancement and speaker identity}
\label{sec:neural}

Speech enhancement is a routine processing step. We applied two pretrained enhancers to the signed clean clips (first 300 clips of the benchmark, PCM16): DNS64~\cite{defossez2020denoiser} and DeepFilterNet3~\cite{schroeter2023deepfilternet3}, each file processed in a fresh invocation. APC verifies $0.827$ of clips after DNS64 at PESQ $3.65$ and $0.973$ after DNS64 with half of the input mixed back ($3.88$), and $0.620$, $0.830$ and $0.870$ after DeepFilterNet3 at its default setting and with its attenuation limited to 12 and 6\,dB ($3.19$--$3.48$; Appendix~\ref{app:neural}). The mark is carried mostly by the magnitude channel under the stronger settings. Signing also keeps the speaker: an ECAPA speaker verifier~\cite{desplanques2020ecapa} matches every signed clip to its source speaker (0 of 300 changed; cosine to the clean clip $0.991$), between AudioSeal ($0.997$) and WavMark ($0.981$) measured with the same model on the same 300 clips.

\subsection{Neural baselines, capacity and coexistence}
\label{sec:sota}
\label{sec:coexist}

Neural watermarks and APC answer different needs. Our reproductions of AudioSeal and WavMark recover their full message under MP3, as published (Appendix~\ref{app:baselines}), but carry 16 and 32 bits, and a neural mark can carry a signature only through a chained layer in which its short message indexes a signed record in a registry. APC verifies offline from the public key. The quality difference between the two is mostly payload. With the original defaults on TED-LIUM ($n{=}30$), APC dialed down to AudioSeal's 16 bits reaches PESQ $4.20$ at $26.4$\,dB SNR against $4.49$ and $27.3$\,dB for AudioSeal, and the full 1{,}160-bit signed payload sits at $15.6$\,dB, so the $72.5\times$ larger payload accounts for about $10.8$\,dB of the SNR difference (Fig.~\ref{fig:capacity} in Appendix~\ref{app:baselines}). Signatures also change who can mint a mark: anyone with the released AudioSeal generator can create valid AudioSeal marks (30/30), whereas no APC mark signed with an attacker key verified (0/30). The two marks can be stacked. With AudioSeal applied first and APC second on the first 300 benchmark clips, AudioSeal still recovers its 16-bit message exactly on 300/300 clips under all eight benchmark attacks, and APC shows no systematic loss against a matched control (changes of $0$ to $-4$ clips of 300).

\subsection{Security analysis}
\label{sec:security}
\label{sec:adversaries}
\label{sec:erasure}

\begin{figure}[t]
  \centering
  \includegraphics[width=\linewidth]{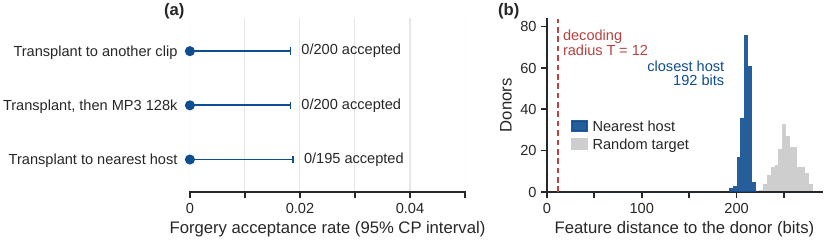}
  \caption{Content-bound mode. (a) Transplant attempts that the verifier accepts: none of 200 naive transplants into unrelated test-clean clips (also after MP3 at 128\,kbps) and none of the 195 executable transplants onto the nearest host in a 5{,}607-clip corpus; exact 95\% intervals. (b) Hamming distance of the 511-bit feature from each donor to a random target and to its nearest eligible host, against the decoding radius $T{=}12$.}
  \label{fig:binding}
\end{figure}

Forgery never succeeded. APC accepted none of the wrong-key verifications of the main benchmark (30{,}480), the ablation arms (76{,}200), the speech envelope (261{,}000) and the music envelope (252{,}648), and none of the corresponding unsigned verifications, and signing with an attacker key never produced an accepted mark. The optional decoder-side search keeps this property: its 1{,}000 unsigned verifications on speech ran about 45 million candidate decodes without an acceptance. The optional content-bound mode rejects payload transplants: it accepted none of 200 naive transplants into unrelated clips (195 of them executable) and none of 195 transplants onto the nearest host found in a 5{,}607-clip corpus, while accepting 8 of 8 duplicates of the same recording, at a PESQ cost of $0.037$ (Fig.~\ref{fig:binding}; Appendix~\ref{app:binding}).

\subsection{Cost}
\label{sec:cost}
On one pinned core of an Intel Xeon Platinum 8480+ (single thread, 10\,s clip, median of 300 timings), the Python reference embeds in $12.35$\,ms and verifies a marked clip in $3.17$\,ms, and the Rust library signs in $2.62$\,ms and verifies in $1.51$\,ms (Appendix~\ref{app:repro}). No GPU is needed.

\section{Discussion and limitations}
\label{sec:discuss}

APC fits capture-time provenance pipelines (C2PA signing, broadcast ingest, archival capture) that keep masters and distribute copies at 96--128\,kbps or more (Section~\ref{sec:envelope}). The absence of learned components is a design choice with measurable benefits: offline verification by a third party with only a public key, deterministic behavior and a single-core cost of milliseconds, for content from any generator. A neural mark for resilience to resynthesis, applied before APC for signing, is the composition we consider most promising.

\paragraph{Limitations.}
\label{sec:limits}
We state the boundaries we measured and the limits we know of.

\paragraph{Resynthesis and desynchronization.} APC is designed to preserve a signature through audio transport. It remains limited under neural codecs, speech resynthesis and processing that changes the time scale. The decoder-side search recovers integer time shifts, but it does not guarantee robustness to speed changes, pitch shifts, reverberation or playback and recapture.

\paragraph{Processing range.} Robustness to noise, requantization, denoising and repeated transcoding depends on the strength and configuration of the processing. Our conclusions are limited to the conditions we evaluated.

\paragraph{Adversarial boundary.} Header-independent verification removes the early rejection that a corrupted length header causes, and APC keeps verifying under the header attacks we tested. Adaptive payload perturbations aimed at the public layout, however, can still make verification fail with small changes in objective quality. This is a limit of erasure resistance, not a signature forgery.

\paragraph{Quality and deployment.} Lowering the embedding strength trades quality against robustness. The platform processing chains were simulated, and performance in real deployments needs further validation.

\section{Conclusion}
\label{sec:conclusion}

APC carries a full Ed25519 signature in the audio waveform through a hybrid phase and magnitude-QIM channel, training-free and verifiable by anyone with the public key, and it proves who signed the embedded message and that the message is intact. It verifies 98.3--99.8\% of 1{,}905 LibriSpeech \texttt{test-clean} clips under eight attacks at an identity PESQ of $3.82$ and 97.1--99.4\% of 2{,}048 clips from 18 corpus sets, accepts none of 544{,}128 wrong-key verifications, survives the platform chains, repeated coding and cropping bounds that its layout predicts, and stacks with AudioSeal. An optional content-bound mode extends the guarantee from the message to the host audio. APC offers asymmetric signing, no training, deterministic behavior and millisecond verification on one CPU core, which makes it a practical provenance layer for audio; Section~\ref{sec:limits} states where its envelope ends.

\label{maintextend}
\FloatBarrier

\begin{ack}
We thank the OASIS Lab at the University of California, Los Angeles for rigorous discussions that sharpened the arguments of this paper.
We thank the Phi Lab Foundation for providing computing server resources.
Fengchen Liu thanks the IT Division at Lawrence Berkeley National Laboratory and Research IT at the University of California, Berkeley for their support.
Funding in direct support of this work: computing resources provided by the Phi Lab Foundation, a nonprofit organization; part of the validation experiments used computing resources at Lawrence Berkeley National Laboratory. This work received no other third-party funding, including no government or university research grants.
Competing interests: The authors declare no competing interests.
\end{ack}

{\small
\bibliographystyle{unsrtnat}
\bibliography{refs/minimal}
}

\appendix
\section{Format and algorithm}
\label{app:format}

The normative description is the format specification (version 1) in the release; this appendix condenses it. Table~\ref{tab:params} lists the parameters, Table~\ref{tab:switches} the switches used by the ablation.

\begin{table}[tbp]
\centering
\caption{Parameters of the APC format; \texttt{wb} is the default profile. Bin $k$ is centered at $k\cdot 44{,}100/2{,}048$\,Hz, and band edges are given at the first and one-past-last bin.}
\label{tab:params}
\footnotesize
\begin{tabular}{@{}l l@{}}
\toprule
Parameter & Value \\
\midrule
Sample rate, STFT & 44.1\,kHz; $N_\text{FFT}{=}2048$, hop 2048, rectangular window, no centering \\
Group & $G{=}8$ frames (16{,}384 samples, 0.37\,s); a 10\,s clip has 215 frames and 26 groups \\
Wideband profile (\texttt{wb}) & phase bins 60--299 (edges 1292--6460\,Hz); magnitude bins 100--339 (2153--7321\,Hz) \\
Narrowband profile (\texttt{nb}) & phase bins 60--299; magnitude bins 16--167 (edges 345--3618\,Hz) \\
Bits per group & phase 240; magnitude 120 (adjacent pairs) \\
QIM step, replica bounds & $\Delta{=}1.0$\,nat; $R_p{=}1$, $R_m{=}5$ \\
Payload & $\text{len}_{16}(M)\,\|\,M\,\|\,S$ (big-endian length), RS over $\mathrm{GF}(2^8)$, 30 parity bytes; $|M|\le159$ \\
Stream per channel & header $\text{len}_{32}(b)$ (big-endian bit count) written three times, then $r$ body copies \\
Layout seed & $\kappa=\mathrm{int}(\text{SHA-256}(K_\text{pub})[0{:}8])\bmod 2^{32}$ (big-endian); MT19937 shuffles \\
Magnitude permutation seed & $(\kappa\oplus\texttt{0xDEADBEEF})\bmod 2^{32}$ \\
Silent-bin erasure & soft value 0 if the magnitudes fall below $\tau{=}10^{-6}$ \\
\bottomrule
\end{tabular}
\end{table}

\begin{table}[tbp]
\centering
\caption{Switches of the reference implementation used by the ablation (Table~\ref{tab:ablation}). Each sets one design element back to its original default.}
\label{tab:switches}
\footnotesize
\begin{tabular}{@{}l l l@{}}
\toprule
Element & APC & Switch (original default) \\
\midrule
Phase write & anchor frame only & \texttt{phase\_frames=8} (all 8 frames of a group) \\
Tail (last partial frame) & copied from the input & \texttt{tail="zero"} (zeroed) \\
Verifier & header value, then lengths 0--159 & \texttt{verifier="header"} (header value only) \\
Silent bins & erased ($\tau{=}10^{-6}$) & \texttt{erasure\_tau=0} (decoded as data) \\
Layout seed & derived from $K_\text{pub}$ & \texttt{seed=}$n$ (fixed integer) \\
\bottomrule
\end{tabular}
\end{table}

\paragraph{Embedding and reading.} Both channels are computed from the same original STFT and combined as $X'=A'\,e^{j\,\mathrm{wrap}(\phi')}$. The phase write sets $\phi'(i_0,k)=\pm\pi/2$ at the anchor frame $i_0=gG$ of group $g$. The magnitude write shifts the two bins of a pair in opposite directions over the 8 frames of a group, as in Section~\ref{sec:mag}, with round-half-even quantization. The reader sums soft values over the body copies, $\sum_r\sigma[96+r|b|+j]$, and thresholds at 0; summing continuous soft values before one threshold, rather than voting on hard bits, avoids the tie bias of majority voting. Conjugate symmetry is preserved by the real-valued inverse transform. Bit $i$ of a stream lives in group $\lfloor i/B\rfloor$ and slot $i\bmod B$, where $B$ is the number of bits per group.

\paragraph{Verification.} For each profile (the given one, or \texttt{wb} then \texttt{nb}) and each channel (phase first), the verifier (1) reads the header by summing its three copies and, if the length fits, decodes the body at that length; otherwise or on failure it (2) tries every message length $m=0,\ldots,159$, decoding the body with the replica count that the length implies. It (3) accepts if RS decoding succeeds, the decoded length prefix is consistent, and Ed25519 verifies $S$ over $M$. The first acceptance wins; nothing else is a positive. An optional resynchronizing search over sample offsets and frame skips (off by default) is used only in the shift experiment of Section~\ref{sec:envelope}.

\begin{figure}[tbp]
  \centering
  \includegraphics[width=\linewidth]{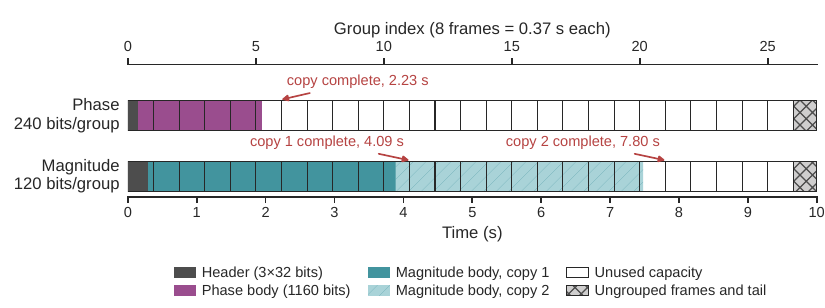}
  \caption{Where the signed stream lives in a 10\,s clip (441{,}000 samples at 44.1\,kHz, 26 groups of 8 frames). Each cell is one group; its fill shows the share of the group's bit slots used (slots are frequency bins, not time). Both channels start with a 96-bit header followed by the 1{,}160-bit body. The phase channel (240 bits per group) holds one copy in groups 0--5; the magnitude channel (120 bits per group) holds $r{=}2$ copies in groups 0--20, since $r=\min(5,\lfloor(3120-96)/1160\rfloor)$. The last 0.34\,s (ungrouped frames and the sample tail) carries no payload.}
  \label{fig:layout}
\end{figure}

\paragraph{Replica counts and minimum duration.} With $C=B\cdot\#\text{groups}$, the body is repeated $r=\max(1,\min(R,\lfloor(C-96)/|b|\rfloor))$ times; signing fails if $C-96<|b|$. For the 1{,}160-bit body, the magnitude channel carries one copy from 4.09\,s, two from 7.80\,s, three from 11.15\,s and five from 18.58\,s; the phase channel needs only 2.23\,s for one copy. The floor therefore depends on the message length: 2.97\,s for an empty message and 6.69\,s for the maximum of 159 bytes, both for the \texttt{wb} profile. A sample-exact sweep on 30 test-clean clips confirmed the 49-byte floor of 180{,}224 samples; above the floor, the phase-channel identity decision matched the 10\,s result on every clip at all 23 durations. Under MP3 at 128\,kbps, verification from 4.1 to 6\,s was 29/30, rising to 30/30 at 8 and 10\,s, where the second magnitude copy is present. Fig.~\ref{fig:layout} shows where the signed stream lives in a 10\,s clip, and Fig.~\ref{fig:spectrogram} the spectrogram of a signed clip and of its difference from the host.

\paragraph{Why Ed25519.} NIST SP~800-131A Rev.~2~\cite{nist_sp800_131Ar2} disallows RSA-1024 signature generation and permits only legacy verification, and C2PA excludes RSA-1024~\cite{c2pa_spec}. Compared with RSA-1024, Ed25519 halves the signature from 128 to 64 bytes, offers about 128-bit rather than below 112-bit classical security~\cite{rfc8032,nist_sp800_131Ar2}, signs faster but verifies single signatures more slowly in the reported benchmark~\cite{bernstein2012ed25519}, and is approved by FIPS 186-5~\cite{nist2023digital}. The shorter payload also shortens the audio needed to carry it. An RSA-PSS backend exists for interoperability; every experiment in this paper uses Ed25519.

\section{Datasets}
\label{app:data}

\paragraph{Main benchmark.} LibriSpeech \texttt{test-clean}~\cite{panayotov2015librispeech} (CC BY 4.0) has 40 speakers, 87 chapters and 2{,}620 utterances. For each speaker and chapter we concatenate the utterances in utterance-id order without gaps, resample from 16 to 44.1\,kHz with soxr at high quality, quantize to 16-bit PCM, and cut consecutive non-overlapping windows of 441{,}000 samples, discarding each chapter's final partial window. The build yields 1{,}905 clips (19{,}050\,s of the 19{,}452.5\,s input; 402.5\,s of chapter tails are discarded). Seventeen clips contain PCM16 clipping (1{,}378 samples in total, at most 254 in one clip). The build is deterministic, and the release lists every clip.

\paragraph{Eighteen corpora.} The corpus evaluation uses 18 sets of ten-second clips drawn from 15 public corpora (FMA, MUSDB18-HQ and AMI contribute two sets each). They cover music, clean read and lecture speech, meeting, broadcast, crowd-sourced and telephone speech, multilingual read speech, environmental recordings and mostly vocoded speech: five music sets (FMA genres 300, GTZAN 240, FMA small 200, MUSDB18-HQ test 138, MUSDB18-HQ 90), two clean speech sets (a separate 150-clip LibriSpeech \texttt{test-clean} set and 30 TED-LIUM clips from 10 talks), and eleven further sets (AMI headset 90, AMI distant microphone 64, AudioSet 90, Common Voice 17 80, DEMAND 72, Earnings-22 90, FLEURS 90, the People's Speech 90, SIMSAMU telephone speech at 8\,kHz 72, VoxConverse 90, WaveFake 72, which is vocoded LJSpeech~\cite{ito2017ljspeech} and in 20 of its 72 clips contains a bona fide LJSpeech recording). All clips are 44.1\,kHz mono PCM16 and exactly 10\,s. The release identifies each clip by source repository, split and row identifier for every set except LibriSpeech, MUSDB18-HQ train and TED-LIUM. The AMI headset and distant-microphone sets cover overlapping meetings: 45 of the 64 distant-microphone windows have a headset counterpart. Table~\ref{tab:asset-licenses} lists the license of every corpus.

\paragraph{Licenses of existing assets.}
Tables~\ref{tab:asset-licenses} and~\ref{tab:model-licenses} list every dataset and pretrained model the evaluation uses, with the version or split, the source and the license stated by its creators, checked on September~29, 2026. Several corpora were obtained from Hugging Face mirrors, and most of these mirrors state no license of their own. In those cases we report the license of the original release, and both sources are named in the table. We use every asset for non-commercial academic evaluation only, and we make no attempt to identify Common Voice speakers. The release contains no audio from, or derived from, any corpus other than LibriSpeech, and its three LibriSpeech excerpts are distributed under CC~BY~4.0 with attribution. The restrictive terms therefore constrain how we use the data, not what we release. They are no derivatives and non-commercial use for TED-LIUM~3, educational use only for MUSDB18-HQ, per-track licenses for FMA, third-party copyright in the VoxConverse audio, and no stated license for the AudioSet audio. GTZAN and the MIT IR Survey state no license, and the creator of GTZAN notes that no copyright permission was obtained for its recordings and asks to be contacted before results are published. We informed him of this use.

\begin{table}[tbp]
  \centering
  \footnotesize
  \setlength{\tabcolsep}{3pt}\renewcommand{\arraystretch}{1.15}\useOriginalUrlSetting 
  \caption{Datasets used in the evaluation. $n$ is the number of 10\,s clips. ``HF'' is a Hugging Face dataset repository; where it is a third-party mirror, the license shown is that of the original release.}
  \label{tab:asset-licenses}
  \begin{tabular}{@{}L{2.3cm}L{2.8cm}L{3.5cm}L{4.7cm}@{}}
    \toprule
    Asset & Version / split used ($n$) & License (as stated by the creators) & Source \\
    \midrule
    LibriSpeech~\cite{panayotov2015librispeech} & \texttt{test-\allowbreak clean}: main benchmark (1{,}905), 150-clip set; \texttt{dev-\allowbreak clean}+\texttt{test-\allowbreak clean}: 5{,}000-clip pool of the baseline reproductions & CC~BY~4.0 & \url{openslr.org/12} \\
    TED-LIUM~3~\cite{hernandez2018tedlium3} & Release~3 dev and test talks, long-form repackaging (30) & CC~BY-NC-ND~3.0; talks are property of TED Conferences LLC & \url{openslr.org/51};\newline HF \texttt{distil-\allowbreak whisper/\allowbreak tedlium-\allowbreak long-\allowbreak form} \\
    MUSDB18-HQ~\cite{rafii2019musdb18hq} & Train (90) and test (138) mixtures & Educational use only; no commercial use without the copyright holders' permission (most tracks under DSD100 terms or CC~BY-NC-SA) & Zenodo 3338373;\newline HF \texttt{danjacobellis/\allowbreak musdb18HQ} \\
    FMA~\cite{defferrard2017fma} & \texttt{fma\_\allowbreak small} (200); genre-labeled FMA repackaging (300) & Audio: license chosen by each artist (Creative Commons family, incl.\ NC and ND); metadata CC~BY~4.0 & \url{github.com/mdeff/fma};\newline HF \texttt{benjamin-\allowbreak paine/\allowbreak free-\allowbreak music-\allowbreak archive-\allowbreak small},\newline \texttt{lewtun/\allowbreak music\_\allowbreak genres} \\
    GTZAN~\cite{tzanetakis2002gtzan} & 10 genres (240) & Not stated; creator notes no copyright permission & \url{marsyas.info};\newline HF \texttt{sanchit-\allowbreak gandhi/\allowbreak gtzan} \\
    AMI~\cite{carletta2006ami} & Test, IHM (90) and SDM (64) & CC~BY~4.0 & \url{groups.inf.ed.ac.uk/ami};\newline HF \texttt{diarizers-\allowbreak community/\allowbreak ami} \\
    FLEURS~\cite{conneau2023fleurs} & Test, 10 locales (90) & CC~BY~4.0 & HF \texttt{google/\allowbreak fleurs} \\
    Common Voice~17.0~\cite{ardila2019commonvoice} & Test, 8 languages (80) & CC0~1.0; users must not attempt to identify speakers & Mozilla;\newline HF \texttt{fixie-\allowbreak ai/\allowbreak common\_\allowbreak voice\_\allowbreak 17\_\allowbreak 0} \\
    People's Speech~\cite{galvez2021peoples} & CC~BY \texttt{dirty} subset, train (90) & CC~BY (2.0--4.0, per source) & HF \texttt{MLCommons/\allowbreak peoples\_\allowbreak speech} \\
    AudioSet~\cite{gemmeke2017audioset} & Balanced, evaluation segments (90) & Labels CC~BY~4.0, ontology CC~BY-SA~4.0; audio from YouTube; no license stated for the audio & \url{research.google.com/audioset};\newline HF \texttt{agkphysics/\allowbreak AudioSet} \\
    DEMAND~\cite{thiemann2013demand} & 18 single-channel files of the mirror (channels 1--16 at 16\,kHz, 1--2 at 48\,kHz), 4 windows each (72) & CC~BY~4.0 (record metadata); CC~BY-SA~3.0 (record text) & Zenodo 1227121;\newline HF \texttt{voice-\allowbreak biomarkers/\allowbreak DEMAND-\allowbreak acoustic-\allowbreak noise} \\
    Earnings-22~\cite{delrio2022earnings} & Test (90) & CC~BY-SA~4.0 (transcripts; audio license not stated separately) & \url{github.com/revdotcom/speech-datasets};\newline HF \texttt{distil-\allowbreak whisper/\allowbreak earnings22} \\
    VoxConverse~\cite{chung2020voxconverse} & Test (90) & CC~BY~4.0, for research purposes; video copyright with its owners & \url{github.com/joonson/voxconverse};\newline HF \texttt{diarizers-\allowbreak community/\allowbreak voxconverse} \\
    SIMSAMU~\cite{simsamu2025} & All dialogs (72) & MIT & HF \texttt{medkit/\allowbreak simsamu},\newline \texttt{diarizers-\allowbreak community/\allowbreak simsamu} \\
    WaveFake~\cite{frank2021wavefake} & LJSpeech-based rows (72; 20 contain a bona fide LJSpeech recording) & CC~BY-SA~4.0; bona fide rows from LJSpeech~\cite{ito2017ljspeech} (public domain) & Zenodo 5642694;\newline HF \texttt{ajaykarthick/\allowbreak wavefake-\allowbreak audio} \\
    MIT IR Survey~\cite{traer2016reverberation} & 270 real room impulse responses (room attacks of the envelope run) & Not stated & \url{mcdermottlab.mit.edu/Reverb/IR_Survey.html} \\
    \bottomrule
  \end{tabular}
\end{table}

\begin{table}[tbp]
  \centering
  \footnotesize
  \setlength{\tabcolsep}{3pt}\renewcommand{\arraystretch}{1.15}\useOriginalUrlSetting 
  \caption{Pretrained models used as attacks and baselines. Hugging Face revisions are those used in the experiments.}
  \label{tab:model-licenses}
  \begin{tabular}{@{}L{2.4cm}L{4.9cm}L{2.5cm}L{3.5cm}@{}}
    \toprule
    Model & Checkpoint (revision) & License & Role \\
    \midrule
    EnCodec~\cite{defossez2023encodec} & \texttt{facebook/\allowbreak encodec\_\allowbreak 24khz} (c1dbe2a) & MIT (code repository; the model card states none) & Neural-codec attack, 1.5--24\,kbps \\
    Vocos~\cite{siuzdak2023vocos} & \texttt{charactr/\allowbreak vocos-\allowbreak mel-\allowbreak 24khz} (0feb3fd) & MIT & Vocoder resynthesis \\
    HiFi-GAN~\cite{kong2020hifigan} (SpeechBrain~\cite{ravanelli2021speechbrain}) & \texttt{speechbrain/\allowbreak tts-\allowbreak hifigan-\allowbreak ljspeech} (17fbdc3) & Apache~2.0 & Vocoder resynthesis \\
    SpeechT5~\cite{ao2022speecht5} & \texttt{microsoft/\allowbreak speecht5\_\allowbreak vc} (c418ba2), \texttt{speecht5\_\allowbreak hifigan} (bb6f429) & MIT & Voice conversion \\
    x-vector (SpeechBrain~\cite{ravanelli2021speechbrain}) & \texttt{speechbrain/\allowbreak spkrec-\allowbreak xvect-\allowbreak voxceleb} (56895a2) & Apache~2.0 & Target-speaker embedding for SpeechT5 \\
    ECAPA-TDNN~\cite{desplanques2020ecapa} (SpeechBrain~\cite{ravanelli2021speechbrain}) & \texttt{speechbrain/\allowbreak spkrec-\allowbreak ecapa-\allowbreak voxceleb} (0f99f2d) & Apache~2.0 & Speaker identity (Section~\ref{sec:neural}) \\
    DNS64~\cite{defossez2020denoiser} & \texttt{facebookresearch/\allowbreak denoiser} 0.1.5, \texttt{dns64} & CC~BY-NC~4.0 (evaluation only) & Speech enhancement \\
    DeepFilterNet3~\cite{schroeter2023deepfilternet3} & release binary and ONNX model v0.5.6 & MIT or Apache~2.0 & Speech enhancement \\
    MetricGAN+~\cite{fu2021metricganplus} (SpeechBrain~\cite{ravanelli2021speechbrain}) & \texttt{speechbrain/\allowbreak metricgan-\allowbreak plus-\allowbreak voicebank} (a196ce2) & Apache~2.0 & Speech enhancement \\
    FreeVC~\cite{li2022freevc}, WavLM~\cite{chen2022wavlm} & FreeVC24, WavLM-Large and speaker encoder (Coqui TTS release \texttt{v0.13.0\_\allowbreak models}) & MIT (FreeVC); MIT (WavLM) & Voice conversion \\
    kNN-VC~\cite{baas2023knnvc} & WavLM-Large with prematched HiFi-GAN vocoder (Coqui TTS release \texttt{v0.25.2\_\allowbreak models}) & MIT & Voice conversion \\
    OpenVoice~V2~\cite{qin2023openvoice} & \texttt{myshell-\allowbreak ai/\allowbreak OpenVoiceV2} converter (f36e7ed) & MIT & Voice conversion \\
    AudioSeal~\cite{sanroman2024audioseal} & \texttt{facebook/\allowbreak audioseal}, 16-bit (3c19eba) & MIT (code and weights) & Neural-watermark baseline; coexistence \\
    WavMark~\cite{chen2023wavmark} & \texttt{M4869/\allowbreak WavMark} (0ad3c7b) & MIT & Neural-watermark baseline \\
    SilentCipher~\cite{singh2024silentcipher} & Not run; published results only & MIT & Cited baseline \\
    \bottomrule
  \end{tabular}
\end{table}

\paragraph{Software.}
The voice-conversion systems run through Coqui TTS~0.27.2 (MPL~2.0), and the models through SpeechBrain~1.1.1 and Transformers~4.55.4 (both Apache~2.0). Codec and signal-processing attacks use an FFmpeg~7.0.2 static build (GPL~3.0 or later as configured). Quality metrics use librosa~1.0.0 (ISC), \texttt{pystoi}~0.4.1 (MIT) and \texttt{pesq}~0.0.4, an MIT-licensed wrapper around a modified copy of the ITU-T P.862 C code. Its intellectual-property notice permits non-commercial evaluation use only. None of these tools is redistributed with our release.

\section{Protocol and metrics}
\label{app:protocol}

\paragraph{Attacks.} MP3 uses \texttt{-b:a 128k} (libmp3lame), OGG uses \texttt{-c:a libvorbis -b:a 128k}, and FLAC uses \texttt{-c:a flac}, all in a static FFmpeg 7.0.2 build. Resampling goes from 44.1 to 16\,kHz and back. The low-pass is a 4th-order Butterworth filter at 8\,kHz applied forward and backward (zero phase). End-cropping zeroes the last 10\% or 20\% of the samples. In the envelope run (Appendix~\ref{app:envelope}), head and middle cropping zero the first or the middle $p$\%, tail removal shortens the clip, and codecs and filters use the same FFmpeg build. Signed audio is quantized to 16-bit PCM before each attack in the primary setting. The message is a fixed 49-byte ASCII string.

\paragraph{Metrics.} Table~\ref{tab:metrics} defines the per-clip metrics. PESQ is wideband at 16\,kHz on the whole clip; the reference is the 16\,kHz resample of the original, and the degraded signal is resampled after trimming both to the common length. STOI is computed at 44.1\,kHz. LSD uses a Hann-window STFT at 44.1\,kHz with $n_\text{fft}{=}2048$ and hop 512, not the coder's frames. Every quality metric compares the attacked watermarked clip with the original, so it includes the embedding distortion.

\paragraph{Statistics.} Verification rates carry 95\% Wilson intervals (exact Clopper--Pearson intervals where stated). Coder comparisons are paired on the same clips: the difference in verification rate carries a paired bootstrap 95\% interval (2{,}000 resamples, seed 42) and an exact McNemar test on the discordant clips; quality differences carry paired bootstrap intervals. Wrong-key controls verify every attacked clip with a second fixed key; unsigned controls verify the unsigned original with the true key.

\begin{table}[tbp]
\centering
\caption{Per-clip evaluation metrics. Better $\uparrow$/$\downarrow$ indicates direction of improvement. $s$ is the original clip and $\tilde{s}$ the attacked watermarked clip (length-aligned); $S_{t,f},\tilde{S}_{t,f}$ are their STFT magnitudes; $(o_i,e_i)$ are the original and extracted bits, $n{=}\min(|\mathbf{o}|,|\mathbf{e}|)$.}
\label{tab:metrics}
\footnotesize
\setlength{\tabcolsep}{4pt}
\begin{tabular}{@{}l l c c@{}}
\toprule
Metric & Definition & Range & Better \\
\midrule
PESQ~\cite{itu_p862_2}    & ITU-T P.862.2 wideband perceptual quality. & $[1.04,\,4.64]$ & $\uparrow$ \\
STOI~\cite{taal2010stoi}  & Short-Time Objective Intelligibility. & $[0,\,1]$ & $\uparrow$ \\
SNR                       & $10\log_{10}\!\big(\|s\|^2/\|s{-}\tilde{s}\|^2\big)$. & dB & $\uparrow$ \\
PSNR                      & $20\log_{10}\!\big(\max|s|/\sqrt{\mathrm{MSE}(s,\tilde{s})}\big)$. & dB & $\uparrow$ \\
LSD                       & $\mathbb{E}_t\!\sqrt{\mathbb{E}_f\!\big(10\log_{10}|S_{t,f}|^2 - 10\log_{10}|\tilde{S}_{t,f}|^2\big)^2}$. & dB & $\downarrow$ \\
BER                       & $\tfrac{1}{n}\sum_{i}\mathbb{1}[e_i\!\neq\! o_i]$ on the RS-encoded payload; capped at $1$ on extraction failure. & $[0,\,1]$ & $\downarrow$ \\
NC                        & Cosine similarity of bipolar codes $(2o_i{-}1)$ and $(2e_i{-}1)$ over $i{=}1,\ldots,n$. & $[-1,\,1]$ & $\uparrow$ \\
Verify\%                  & Fraction of clips for which RS decoding succeeds \emph{and} Ed25519 verifies. & $[0,\,100]\%$ & $\uparrow$ \\
\bottomrule
\end{tabular}
\end{table}

\section{Main benchmark detail}
\label{app:legacy}
\label{app:benchmark}

\begin{figure}[tbp]
  \centering
  \includegraphics[width=\linewidth]{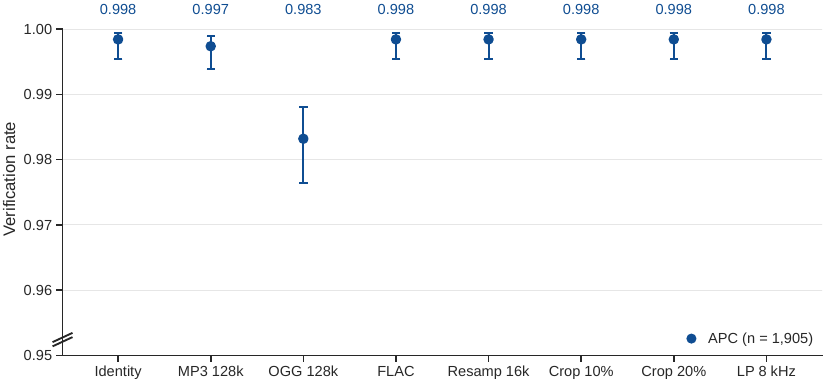}
  \caption{APC verification rate per attack on the 1{,}905 test-clean clips of Table~\ref{tab:results} (one fixed key pair, PCM16 output), with 95\% Wilson intervals. Wrong-key and unsigned false accepts are 0 under every attack.}
  \label{fig:success}
\end{figure}

\begin{figure}[tbp]
  \centering
  \includegraphics[width=\linewidth]{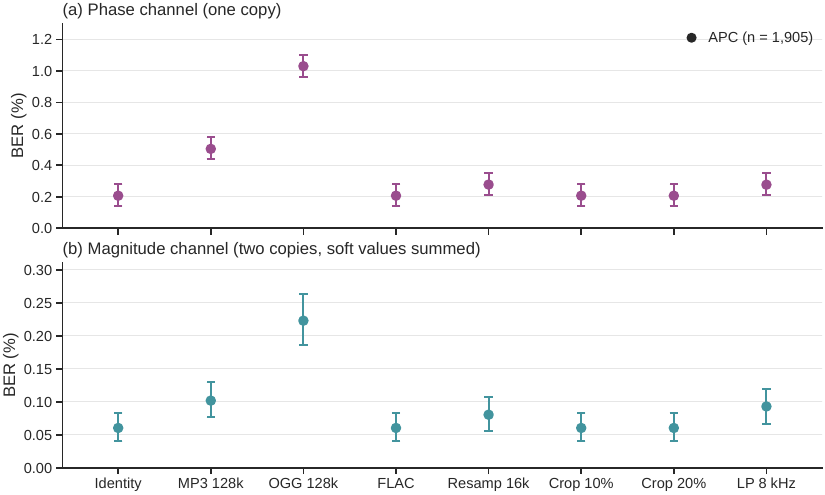}
  \caption{Bit-error rate of the Reed--Solomon-encoded body before decoding, per channel and attack, on the 1{,}905 clips of Table~\ref{tab:results}. (a) Phase channel, one copy. (b) Magnitude channel, soft values of the two copies summed before the sign decision. Points are means over clips; error bars are 95\% bootstrap intervals of the mean. Crop 10\% and 20\% zero the tail after the last payload bit, so they equal identity.}
  \label{fig:ber}
\end{figure}

Figures~\ref{fig:success} and~\ref{fig:ber} give, for the main benchmark of Table~\ref{tab:results}, the verification rate per attack with its interval and the bit-error rate of each channel before decoding.

\section{Quality}
\label{app:quality}

\begin{figure}[tbp]
  \centering
  \includegraphics[width=\linewidth]{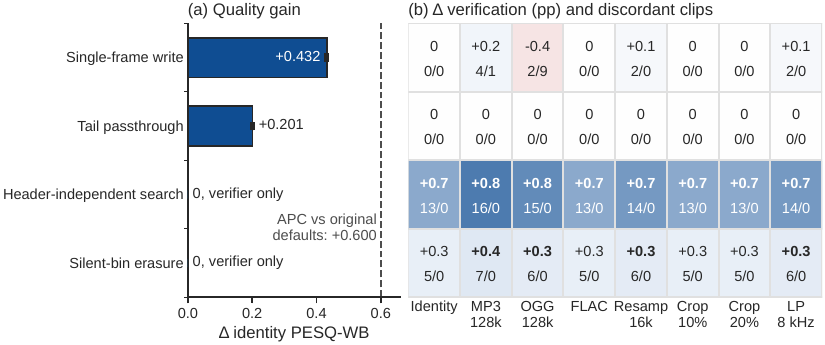}
  \caption{Effect of switching one design element back to its original default on the 1{,}905 test-clean clips (same key pair and clips for all arms, PCM16 output, header-independent verifier unless switched). Every value is APC minus the ablated arm. (a) Paired identity PESQ difference with bootstrap 95\% intervals; the dashed line is the difference to all four original defaults at once ($+0.600$). The single-frame write and the tail passthrough account for $+0.432$ and $+0.201$, which exceeds the total, so the contributions are not additive. The length search and the silent-bin erasure act only in the verifier and leave the signed audio unchanged. (b) Verification difference in percentage points per attack, with the number of clips that verify only under APC or only under the ablated arm; bold marks an exact McNemar $p<0.05$. The single-frame write is not neutral for verification: it gains on MP3 and loses 7 net clips on OGG 128\,k ($p{=}0.065$).}
  \label{fig:ablation}
\end{figure}

\begin{figure}[tbp]
  \centering
  \includegraphics[width=\linewidth]{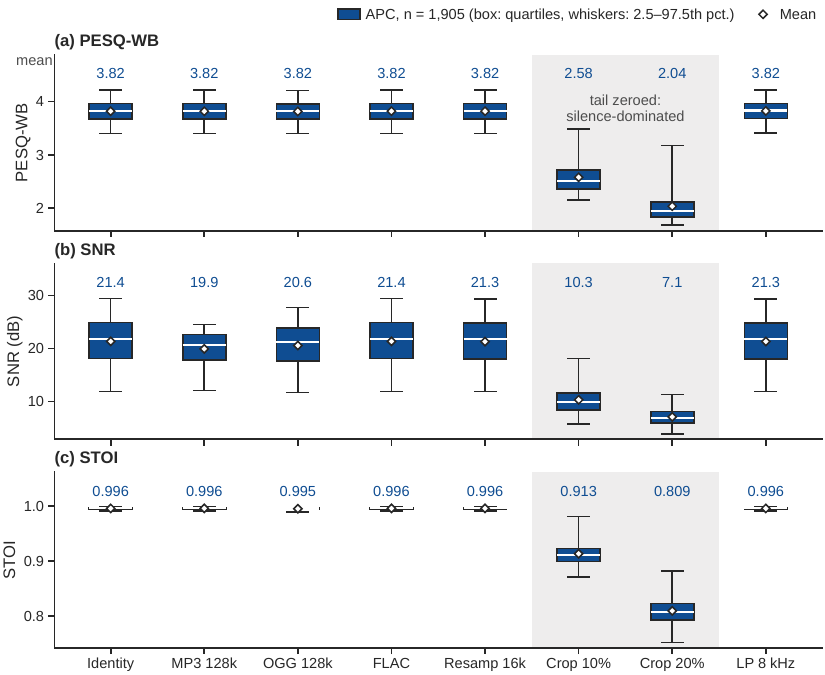}
  \caption{Quality of the attacked signal against the original clip for APC on the 1{,}905 test-clean clips (PCM16 output): PESQ-WB, SNR and STOI per attack. Boxes span the quartiles, whiskers the 2.5th to 97.5th percentiles, lines mark the median and diamonds the mean. The crop attacks replace the last 10\% or 20\% of the clip with silence, so their scores are dominated by the silenced segment and not by the watermark.}
  \label{fig:quality}
\end{figure}

\paragraph{Replication does not repair a gated header.} In a replica-placement run on 150 LibriSpeech \texttt{test-clean} clips with the original defaults, a header-trusting verifier gave $0.000$ at every head-crop depth from 5\% to 50\% for replica counts from 1 to 8, while five replicas cost up to $1.23$ PESQ ($3.13\rightarrow1.90$). A header-independent verifier with the single-frame write and one replica restored head-crop verification to $1.000$ at 5\% and $0.993$ at 10--40\%. The main-benchmark numbers of APC are in Tables~\ref{tab:results} and~\ref{tab:ablation} and in Figs.~\ref{fig:ablation} and~\ref{fig:quality}; head cropping on 1{,}000 clips is in Table~\ref{tab:envelope} and Fig.~\ref{fig:headcrop}.

\subsection{Operating points}
\label{app:oppoint}

\begin{figure}[tbp]
  \centering
  \includegraphics[width=\linewidth]{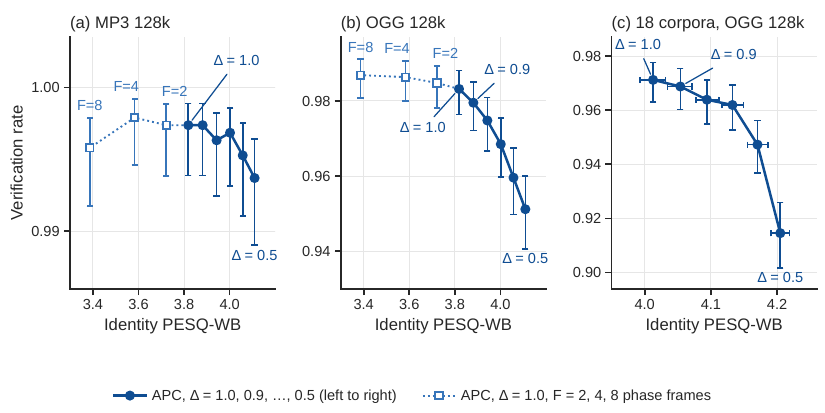}
  \caption{Quality against robustness when the QIM step $\Delta$ of the magnitude channel is reduced, with the verifier configured for the same $\Delta$. Each point shows mean identity PESQ-WB ($x$, bootstrap 95\% interval) and the verification rate after MP3 or OGG at 128\,kbps ($y$, 95\% Wilson interval). (a, b) LibriSpeech \texttt{test-clean} ($n{=}1{,}905$ clips of 10\,s); (c) the 18 corpus sets ($n{=}2{,}048$; PESQ is undefined for 52 clips). Blue: APC from $\Delta{=}1.0$ (default, left) to $0.5$ (right) in steps of 0.1; light blue: APC at $\Delta{=}1.0$ with 2, 4 or 8 phase frames. On LibriSpeech, PESQ rises from 3.82 to 4.11 as $\Delta$ decreases.}
  \label{fig:oppoint}
\end{figure}

Table~\ref{tab:oppoint} gives a tunable operating-point curve, measured on APC with the 1{,}905 clips of Table~\ref{tab:results} (PCM16, one fixed key pair; the verifier is given the same $\Delta$). Lowering the QIM step $\Delta$ raises identity PESQ from $3.82$ to $4.11$ on test-clean and from $4.01$ to $4.20$ on the 2{,}048 corpus clips. At $\Delta{=}0.9$ PESQ rises by $0.06$ with no significant change in verification (OGG $p{=}0.14$), and MP3 stays within 0.4 points of $\Delta{=}1.0$ at every step on test-clean. No point accepted a wrong-key or unsigned clip. $\Delta$ thus gives the signer a quality setting, with the verification at each setting listed in Table~\ref{tab:oppoint}.
Fig.~\ref{fig:oppoint} shows the same points with their intervals.

\begin{table}[tbp]
\centering
\caption{Operating points along the QIM step $\Delta$ (PCM16; the verifier is given the same $\Delta$). APC on the 1{,}905 \texttt{test-clean} clips of Table~\ref{tab:results} and on the 2{,}048 corpus clips of Section~\ref{sec:corpora} (corpus PESQ over the 1{,}996 clips with a valid PESQ). $\Delta$OGG is the paired difference from $\Delta{=}1.0$ with the exact McNemar $p$.}
\label{tab:oppoint}
\footnotesize
\setlength{\tabcolsep}{4pt}
\IfFileExists{generated/tab_oppoint.tex}{\begin{tabular}{@{}lccccccc@{}}
\toprule
 & \multicolumn{4}{c}{LibriSpeech \texttt{test-clean} ($n{=}1{,}905$)} & \multicolumn{3}{c}{18 corpora ($n{=}2{,}048$)} \\
\cmidrule(lr){2-5}\cmidrule(l){6-8}
$\Delta$ & PESQ & MP3 & OGG & $\Delta$OGG, pp ($p$) & PESQ & MP3 & OGG \\
\midrule
1.0 & 3.82 & 0.997 & 0.983 & -- & 4.01 & 0.985 & 0.971 \\
0.9 & 3.88 & 0.997 & 0.980 & \ensuremath{-}0.4 (0.14) & 4.05 & 0.984 & 0.969 \\
0.8 & 3.94 & 0.996 & 0.975 & \ensuremath{-}0.8 (0.0015) & 4.09 & 0.983 & 0.964 \\
0.7 & 4.00 & 0.997 & 0.969 & \ensuremath{-}1.5 (\ensuremath{7.7\times10^{-7}}) & 4.13 & 0.979 & 0.962 \\
0.6 & 4.06 & 0.995 & 0.960 & \ensuremath{-}2.4 (\ensuremath{6.8\times10^{-13}}) & 4.17 & 0.974 & 0.947 \\
0.5 & 4.11 & 0.994 & 0.951 & \ensuremath{-}3.2 (\ensuremath{6.8\times10^{-16}}) & 4.20 & 0.968 & 0.915 \\
\bottomrule
\end{tabular}
}{\textbf{[generated/tab\_oppoint.tex missing]}}
\end{table}

\section{Corpora detail}
\label{app:corpora}

\begin{figure}[tbp]
  \centering
  \includegraphics[width=\linewidth]{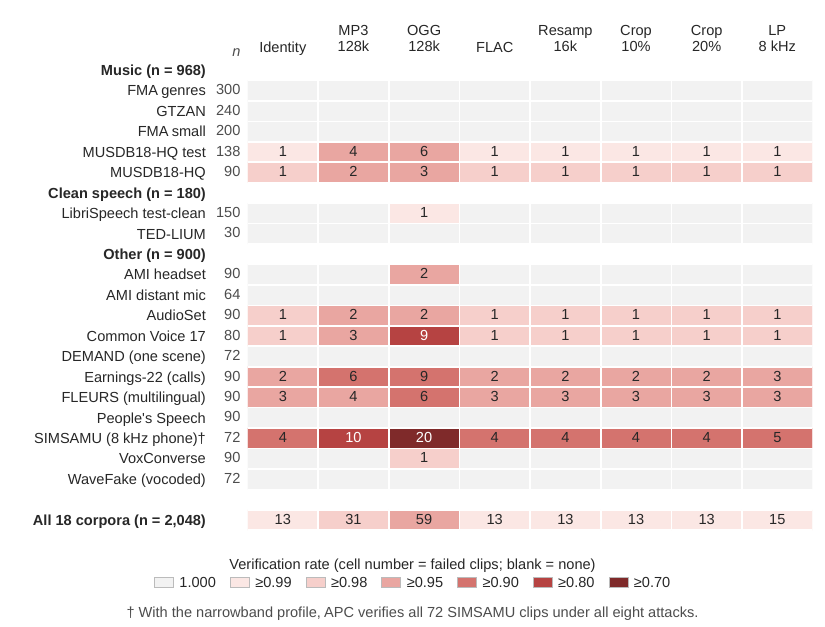}
  \caption{APC verification per corpus and attack on 2{,}048 clips of 10\,s from 18 sets of 15 public corpora (968 music, 180 clean speech, 900 other; wideband profile, one fixed key pair, PCM16 output). Shading bins the verification rate; the number in a cell is the count of clips that fail to verify (blank means none). Wrong-key and unsigned false accepts are 0 in every cell. With the narrowband profile, the 8\,kHz telephone corpus (SIMSAMU) verifies 72/72 under every attack. WaveFake is vocoded LJSpeech; 20 of its 72 clips contain a bona fide LJSpeech recording.}
  \label{fig:corpora}
\end{figure}

Table~\ref{tab:corpora} reports every corpus of Section~\ref{sec:corpora} under identity, MP3 and OGG; Fig.~\ref{fig:corpora} shows all eight attacks.

\paragraph{Narrowband telephone profile.} The wideband magnitude band (2153--7321\,Hz) carries only 4.5\% of the energy of an 8\,kHz SIMSAMU call, whereas the narrowband band (345--3618\,Hz) carries 94.4\%. The signer therefore selects the narrowband profile for narrowband capture, and the verifier tries both profiles. On the 72 SIMSAMU clips the narrowband profile verifies $1.000$ under all eight attacks, at a paired identity PESQ cost of $0.25$ relative to the wideband profile. On wideband content the wideband band already verifies, so the low band would spend quality for no gain.

\begin{table}[tbp]
\centering
\caption{APC verification per corpus (identity, MP3 128\,kbps and OGG 128\,kbps; PCM16 after signing). Group rows aggregate their corpora.}
\label{tab:corpora}
\footnotesize
\setlength{\tabcolsep}{3.5pt}
\IfFileExists{generated/tab_corpora.tex}{\begin{tabular}{@{}lrccc@{}}
\toprule
Corpus & $n$ & Identity & MP3 128\,kbps & OGG 128\,kbps \\
\midrule
FMA (genres) & 300 & 1.000 & 1.000 & 1.000 \\
FMA (small) & 200 & 1.000 & 1.000 & 1.000 \\
GTZAN & 240 & 1.000 & 1.000 & 1.000 \\
MUSDB18-HQ & 90 & 0.989 & 0.978 & 0.967 \\
MUSDB18-HQ (test) & 138 & 0.993 & 0.971 & 0.957 \\
\textit{Music} & \textit{968} & \textit{0.998} & \textit{0.994} & \textit{0.991} \\
\midrule
LibriSpeech \texttt{test-clean} & 150 & 1.000 & 1.000 & 0.993 \\
TED-LIUM 3 & 30 & 1.000 & 1.000 & 1.000 \\
\textit{Clean speech} & \textit{180} & \textit{1.000} & \textit{1.000} & \textit{0.994} \\
\midrule
AMI (headset) & 90 & 1.000 & 1.000 & 0.978 \\
AMI (distant) & 64 & 1.000 & 1.000 & 1.000 \\
AudioSet & 90 & 0.989 & 0.978 & 0.978 \\
Common Voice 17 & 80 & 0.988 & 0.963 & 0.888 \\
DEMAND (one scene) & 72 & 1.000 & 1.000 & 1.000 \\
Earnings-22 & 90 & 0.978 & 0.933 & 0.900 \\
FLEURS & 90 & 0.967 & 0.956 & 0.933 \\
People's Speech & 90 & 1.000 & 1.000 & 1.000 \\
SIMSAMU (8\,kHz) & 72 & 0.944 & 0.861 & 0.722 \\
VoxConverse & 90 & 1.000 & 1.000 & 0.989 \\
WaveFake & 72 & 1.000 & 1.000 & 1.000 \\
\textit{Further corpora} & \textit{900} & \textit{0.988} & \textit{0.972} & \textit{0.946} \\
\midrule
\textbf{All} & \textbf{2{,}048} & \textbf{0.994} & \textbf{0.985} & \textbf{0.971} \\
SIMSAMU, narrowband profile & 72 & 1.000 & 1.000 & 1.000 \\
\bottomrule
\end{tabular}
}{\textbf{[generated/tab\_corpora.tex missing]}}
\end{table}

\section{Transport envelope detail}
\label{app:envelope}

\paragraph{Runs.} APC was run on the first 1{,}000 clips of a fixed seeded shuffle of the 1{,}905 benchmark clips and on the 968 music clips of Section~\ref{sec:corpora}, with the keys, message and PCM16 quantization of Table~\ref{tab:results}, under 261 attack configurations in eight families, each family with its own identity control. The shift family also ran APC with the optional decoder-side resynchronization of the format specification, which scores 2{,}048 sample offsets with up to 10 frame skips and decodes the best 8 offsets. No wrong-key verification (261{,}000 on speech, 252{,}648 on music) and no unsigned verification was accepted. The 1{,}000 unsigned verifications of the resynchronizing arm on speech ran the full search, with about 45 million candidate decodes and 98{,}512 signature checks. Filters and codecs use FFmpeg~7.0.2.

\begin{table}[tbp]
\centering
\caption{Transport conditions beyond the eight benchmark attacks: APC verification rate on 1{,}000 LibriSpeech \texttt{test-clean} speech clips and 968 music clips (PCM16 after signing). Ranges give the minimum and maximum over the listed settings. The paragraphs below give every level; no wrong-key or unsigned verification was accepted.}
\label{tab:envelope}
\footnotesize
\setlength{\tabcolsep}{3pt}
\IfFileExists{generated/tab_envelope_speech_music.tex}{\begin{tabular}{@{}lcc@{}}
\toprule
Condition & Speech ($n{=}1{,}000$) & Music ($n{=}968$) \\
\midrule
\multicolumn{3}{@{}l}{\textit{Crop and delay}} \\
Identity & 0.997 & 0.998 \\
Tail zeroed, 10--84\% & 0.972--0.997 & 0.979--0.998 \\
Tail removed, 10--70\% & 0.980--0.997 & 0.989--0.998 \\
Head zeroed, 5 / 20 / 40\% & 0.992 / 0.970 / 0.923 & 0.993 / 0.988 / 0.978 \\
Middle zeroed, 10--50\% & 0.980--0.991 & 0.989--0.992 \\
Delay of 1 sample & 0.983 & 0.992 \\
Delay of 1 frame (2{,}048 samples) & 0.930 & 0.979 \\
\midrule
\multicolumn{3}{@{}l}{\textit{Decoder-side search}} \\
26 delays, 1--16{,}384 samples & 0.976--0.997 & 0.991--0.998 \\
Advance of 44 samples & 0.859 & 0.962 \\
Raw ADTS AAC, 128\,kbps & 0.976 & 0.988 \\
\midrule
\multicolumn{3}{@{}l}{\textit{Signal processing}} \\
Resampling round trip, 16--48\,kHz & 0.997 & 0.998 \\
Low-pass, 7\,kHz & 0.991 & 0.995 \\
Brown noise, 20 / 10\,dB SNR & 0.983 / 0.935 & 0.997 / 0.994 \\
Requantization to 12 bits & 0.954 & 0.996 \\
Packet loss, 1 / 3 / 5\% & 0.997 / 0.991 / 0.960 & 0.998 / 0.993 / 0.980 \\
Loudness normalization, $-14$ / $-23$\,LUFS & 0.997 / 0.997 & 0.998 / 0.998 \\
Dynamic normalization (dynaudnorm) & 0.997 & 0.998 \\
Speech normalization (speechnorm) & 0.997 & 0.998 \\
Compressor (4:1) & 0.995 & 0.996 \\
Equalizer, $+6$\,dB at 3\,kHz & 0.968 & 0.973 \\
Volume $\times 0.5$ / $\times 2$ & 0.986 / 0.997 & 0.997 / 0.932 \\
\midrule
\multicolumn{3}{@{}l}{\textit{Codecs}} \\
MP3, 96--256\,kbps & 0.984--0.997 & 0.989--0.998 \\
Opus, 96--128\,kbps & 0.963--0.989 & 0.987--0.994 \\
AAC (MP4), 128--192\,kbps & 0.988--0.996 & 0.991--0.994 \\
OGG Vorbis, 128--192\,kbps & 0.978--0.997 & 0.991--0.998 \\
\midrule
\multicolumn{3}{@{}l}{\textit{Chains}} \\
Upload chain & 0.995 & 0.994 \\
Podcast chain & 0.991 & 0.994 \\
Broadcast chain & 0.985 & 0.991 \\
Archive chain & 0.998 & 0.998 \\
2--5 benchmark attacks in series & 0.953--0.964 & 0.987--0.989 \\
MP3 128\,kbps, 5 generations & 0.979 & 0.975 \\
OGG 128\,kbps, 3 generations & 0.954 & 0.980 \\
\bottomrule
\end{tabular}
}{\textbf{[generated/tab\_envelope\_speech\_music.tex missing]}}
\end{table}

\begin{table}[tbp]
\centering
\caption{Lossy codecs from 96\,kbps upward: APC verification rate on 1{,}000 LibriSpeech \texttt{test-clean} clips and 968 music clips, by bitrate (kbps). AAC is in an MP4 container. A dash means not run.}
\label{tab:codecs}
\footnotesize
\setlength{\tabcolsep}{4pt}
\IfFileExists{generated/tab_codecs_fav.tex}{\begin{tabular}{@{}llcccc@{}}
\toprule
Codec & Material & 96\,kbps & 128\,kbps & 192\,kbps & 256\,kbps \\
\cmidrule(l){3-6}
MP3 & Speech & 0.984 & 0.995 & 0.997 & 0.997 \\
 & Music & 0.989 & 0.994 & 0.997 & 0.998 \\
\addlinespace
Opus & Speech & 0.963 & 0.989 & -- & -- \\
 & Music & 0.987 & 0.994 & -- & -- \\
\midrule
Codec & Material & 128\,kbps & 192\,kbps & & \\
\cmidrule(l){3-4}
AAC (MP4) & Speech & 0.988 & 0.996 & & \\
 & Music & 0.991 & 0.994 & & \\
\addlinespace
OGG Vorbis & Speech & 0.978 & 0.997 & & \\
 & Music & 0.991 & 0.998 & & \\
\bottomrule
\end{tabular}
}{\textbf{[generated/tab\_codecs\_fav.tex missing]}}
\end{table}

\begin{figure}[tbp]
  \centering
  \begin{minipage}[t]{0.49\linewidth}
  \centering
  \includegraphics[width=\linewidth]{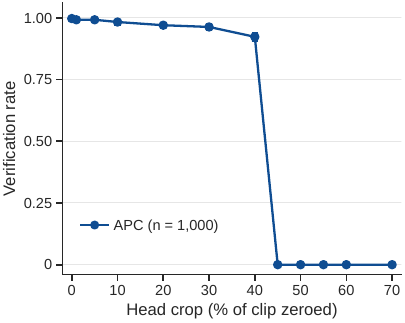}
  \caption{Head cropping on 1{,}000 LibriSpeech \texttt{test-clean} clips (PCM16): the first $p$\% of each clip is set to zero; 95\% Wilson intervals. Because the verifier does not need the stream header at the start of the clip, APC verifies $0.992$ to $0.923$ at 5--40\%, every acceptance through the magnitude channel and the length search.}
  \label{fig:headcrop}
  \end{minipage}\hfill
  \begin{minipage}[t]{0.49\linewidth}
  \centering
  \includegraphics[width=\linewidth]{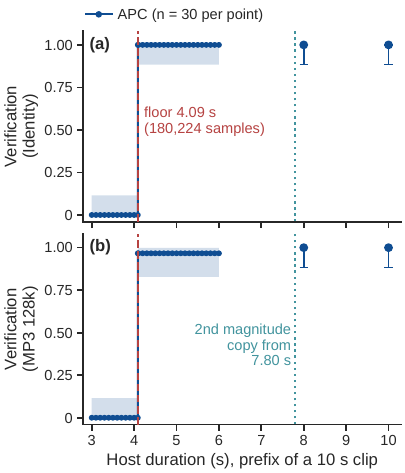}
  \caption{Verification against host duration ($n{=}30$ \texttt{test-clean} clips, each shortened to its first 3--10\,s). The dashed line is the closed-form floor of 180{,}224 samples (4.09\,s); under MP3 the step between 6 and 8\,s matches the second magnitude copy, predicted at 7.80\,s.}
  \label{fig:duration}
  \end{minipage}
\end{figure}

\paragraph{Cropping.} Zeroing the tail keeps APC at $0.987$--$0.997$ up to 60\%, at $0.980$ from 70\% to 80\%, $0.977$ at 81\% and $0.972$ from 82\% to 84\% (Fig.~\ref{fig:bounds}). The layout predicts the edge: the phase copy ends in group 5, whose read frame covers samples 81{,}920--83{,}967, so a tail crop of up to $1-83{,}968/441{,}000=81.0\%$ leaves every phase bit intact, and RS absorbs the loss of the 56 bits (7 bytes) in that group, which extends the bound to $1-67{,}584/441{,}000=84.7\%$. Removing the tail instead, which shortens the file, keeps $0.987$--$0.997$ up to 50\% and $0.980$ at 70\%; one phase copy needs 98{,}304 samples (6 groups). Zeroing the head keeps $0.992$, $0.983$, $0.970$, $0.963$ and $0.923$ at 5, 10, 20, 30 and 40\%, every acceptance through the magnitude channel and the length search (Fig.~\ref{fig:headcrop}), and zeroing the middle 10, 30 and 50\% of the clip keeps $0.991$, $0.985$ and $0.980$. Music behaves the same way: $0.979$ with 84\% of the tail zeroed and $0.978$--$0.993$ with 5--40\% of the head zeroed. The closed-form duration floor of Section~\ref{sec:extract} is confirmed to the sample (Fig.~\ref{fig:duration}).

\begin{figure}[tbp]
  \centering
  \includegraphics[width=\linewidth]{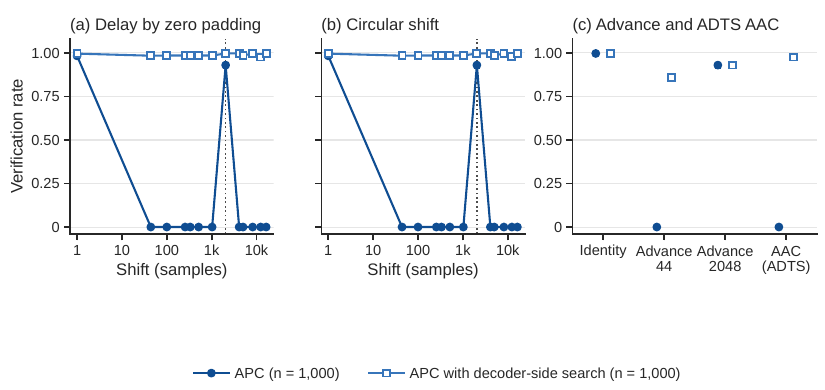}
  \caption{Integer time shifts on 1{,}000 LibriSpeech \texttt{test-clean} clips (PCM16) for APC without and with its optional decoder-side resynchronization (a search over sample offsets and frame skips; the signed audio is unchanged). (a) Delay by prepending zeros and (b) circular shift, at 13 delays from 1 to 16{,}384 samples; the dotted line marks one STFT frame. With the search all 26 settings verify at $0.976$--$0.997$. (c) Advancing the clip by 44 samples ($0.859$ with the search) and by one frame, and AAC at 128\,kbps in a raw ADTS stream ($0.976$ with the search). Error bars are 95\% Wilson intervals.}
  \label{fig:resync}
\end{figure}

\paragraph{Shifts.} Without the search, a delay of one sample ($0.983$) and of exactly one frame of 2{,}048 samples ($0.930$) verify, zero-padded or circular, because the magnitude reader averages each bit over the 8 frames of its group. With the search, APC verifies $0.976$--$0.997$ at all 26 tested shifts from 1 to 16{,}384 samples, $0.859$ after a 44-sample advance (the first samples removed and the end padded) and $0.930$ after a 2{,}048-sample advance, and raw ADTS AAC at 128\,kbps, whose encoder delay shifts the signal, verifies at $0.976$ (Fig.~\ref{fig:resync}). On music the search recovers every tested shift at $0.991$--$0.998$ and raw ADTS AAC at $0.988$. Round trips through 16--48\,kHz keep APC at $0.997$, and a 7\,kHz low-pass at $0.991$.

\paragraph{Codecs, processing and chains.} Table~\ref{tab:codecs} gives the codec rates from 96\,kbps upward, where MP3 and Opus verify $0.963$ or more on speech and AAC and OGG from 128\,kbps; FLAC keeps $0.997$. Loudness normalization to $-14$ and $-23$\,LUFS, dynamic normalization, speech normalization and gain changes keep $0.986$--$0.997$, a compressor $0.995$ and a $+6$\,dB equalizer at 3\,kHz $0.968$. Upload (MP3, $-14$\,LUFS, 48\,kHz round trip), podcast (8\,kHz low-pass, MP3, 10\% tail crop), broadcast ($-23$\,LUFS, AAC, 15\,kHz low-pass) and archive (FLAC, 48\,kHz round trip, FLAC) chains keep $0.995$, $0.991$, $0.985$ and $0.998$, chains of two to five benchmark attacks $0.953$--$0.964$, MP3 followed by AAC at 96\,kbps $0.927$, and two, three and five generations of MP3 at 128\,kbps $0.992$, $0.988$ and $0.979$ (AAC: $0.978$, $0.960$, $0.949$; OGG: $0.958$ and $0.954$ after two and three). Brown noise keeps $0.983$--$0.998$ from 20 to 50\,dB SNR and $0.935$ at 10\,dB, 12-bit requantization $0.954$, and packet loss of 1, 3 and 5\% (20\,ms packets) $0.997$, $0.991$ and $0.960$.

\section{Neural speech enhancement and speaker identity}
\label{app:neural}

\begin{figure}[tbp]
  \centering
  \begin{minipage}[t]{0.49\linewidth}
  \centering
  \includegraphics[width=\linewidth]{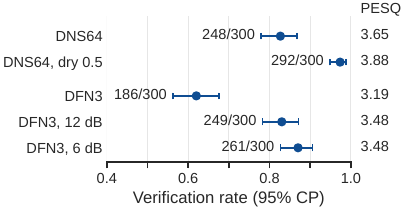}
  \caption{APC verification after neural speech enhancement of signed clean speech ($n{=}300$), with 95\% Clopper--Pearson intervals, against the PESQ-WB of the enhanced signal.}
  \label{fig:enhance}
  \end{minipage}\hfill
  \begin{minipage}[t]{0.49\linewidth}
  \centering
  \includegraphics[width=\linewidth]{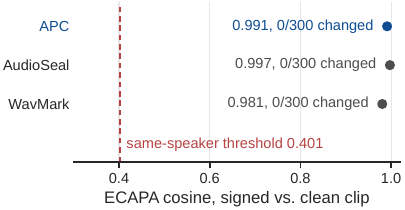}
  \caption{ECAPA cosine similarity between each signed clip and its clean original. Means with bootstrap 95\% intervals on the same 300 clips; APC lies between the AudioSeal and WavMark references, and no signed clip falls below the speaker-verification threshold (0.401).}
  \label{fig:speaker}
  \end{minipage}
\end{figure}

\paragraph{Speech enhancement.} The two enhancers were applied to the signed PCM16 files of the first 300 benchmark clips, without added noise, so that the measurement isolates what enhancement does to a clean signed recording. DNS64~\cite{defossez2020denoiser} is the causal waveform-domain model of the \texttt{denoiser} package, run at its default setting and with a dry mix of 0.5 (half of the input mixed back). DeepFilterNet3~\cite{schroeter2023deepfilternet3} is the official release binary with its ONNX model, run at its default setting and with its attenuation limited to 12 and 6\,dB; every file was processed in a separate invocation. Signing, verification and the wrong-key and unsigned controls are those of Table~\ref{tab:results}; no control was accepted. Table~\ref{tab:enhance} and Fig.~\ref{fig:enhance} give the results. Under the stronger settings the mark is carried mostly by the magnitude channel; under the milder settings the phase channel carries an increasing share (DeepFilterNet3 limited to 6\,dB: phase $0.693$, magnitude $0.177$ of clips).

\begin{table}[tbp]
\centering
\caption{Neural speech enhancement of signed clean speech (first 300 benchmark clips, PCM16): APC verification with exact 95\% Clopper--Pearson intervals and PESQ-WB of the enhanced signal against the original clip.}
\label{tab:enhance}
\footnotesize
\setlength{\tabcolsep}{4pt}
\IfFileExists{generated/tab_enhance.tex}{\begin{tabular}{@{}lccc@{}}
\toprule
Enhancer, setting & Verified (of 300) & Rate [95\% CP] & PESQ \\
\midrule
Signed clip, no enhancement & 299 & 0.997 [0.982, 1.000] & 3.81 \\
\midrule
DNS64, default & 248 & 0.827 [0.779, 0.868] & 3.65 \\
DNS64, dry 0.5 (half input mixed back) & 292 & 0.973 [0.948, 0.988] & 3.88 \\
DeepFilterNet3, default & 186 & 0.620 [0.562, 0.675] & 3.19 \\
DeepFilterNet3, attenuation limit 12\,dB & 249 & 0.830 [0.783, 0.871] & 3.48 \\
DeepFilterNet3, attenuation limit 6\,dB & 261 & 0.870 [0.827, 0.906] & 3.48 \\
\bottomrule
\end{tabular}
}{\textbf{[generated/tab\_enhance.tex missing]}}
\end{table}

\paragraph{Speaker identity.} We embedded every clean and signed clip with the ECAPA-TDNN speaker model~\cite{desplanques2020ecapa} at 16\,kHz and calibrated a same-speaker threshold on all pairs of the 1{,}905 clean clips (equal error rate 0.3\%, threshold 0.401). Every APC-signed clip stays above the threshold against its own clean clip (0 of 300 changed; mean cosine 0.991), and a leave-one-out centroid test assigns all 300 signed clips to their source speaker among the 40 test-clean speakers. The two neural watermarks, embedded with their released models, give mean cosines of 0.997 (AudioSeal) and 0.981 (WavMark) on the same 300 clips, and no clip below the threshold on all 1{,}905 clips (Fig.~\ref{fig:speaker}).

\section{Security analysis detail}
\label{app:adv}

The erasure and header analyses below were measured with the original defaults of Table~\ref{tab:ablation} (eight-frame phase write, header-trusting verifier) on smaller sets; each is labeled with its configuration.

\begin{figure}[tbp]
\centering
\includegraphics[width=\linewidth]{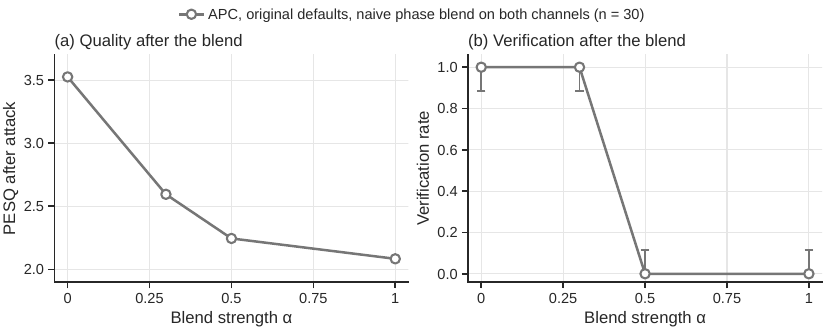}
\caption{Naive white-box erasure of APC with the original defaults (TED-LIUM, $n{=}30$): (a) mean PESQ after the attack and (b) verification rate as the attacker blends the phase at the public bin set toward uniform random phase with strength $\alpha$, in both channels. Error bars are 95\% Clopper--Pearson intervals.}
\label{fig:erasure}
\end{figure}

\paragraph{Naive white-box erasure.} The attacker knows the public layout, reconstructs the watermarked bins exactly, and blends random phase into those bins only, $\phi'(i,k)=\mathrm{wrap}\big((1-\alpha)\phi(i,k)+\alpha u(i,k)\big)$ with $u\sim\mathrm{Unif}(-\pi,\pi]$, applied to both channels. On 30 TED-LIUM clips with the original defaults, the mark survives a blend of $\alpha{=}0.3$, which already costs $0.93$ PESQ, and erasing it needs $\alpha{=}0.5$, which costs $1.28$ PESQ (Fig.~\ref{fig:erasure}). A metadata-only signature, by contrast, costs the attacker no distortion at all to strip.

\paragraph{Header gate.} The stream header lies outside the RS parity, which is why APC's verifier does not depend on it. On 30 TED-LIUM clips (10 talks), changing three, six or twelve phase bins and as many magnitude bin-groups of the header drives a header-trusting verifier to $0.000$ with PESQ unchanged to three decimals. A header-independent verifier, a blind length search, verified all 100 of 100 LibriSpeech clips under a two-bit header attack. APC's verifier closes the gate by construction: it tries the header value and otherwise every admissible length, and accepts only a payload whose RS decoding succeeds and whose signature verifies.

\paragraph{False-accept controls.} A blind length search with the original defaults (lengths 512--4{,}096 bits in steps of 8) over 100 LibriSpeech clips under a wrong key and on unmarked audio produced 0 false accepts in 152{,}800 candidate decodes, of which 295 reached the Ed25519 check. On the main benchmark (Table~\ref{tab:results}) no wrong-key or unsigned verification was accepted. No wrong-key candidate passed RS decoding, because the key-derived layout differs, and unsigned clips reached the signature check 3{,}866 times. The ablation arms of Table~\ref{tab:ablation} add 76{,}200 wrong-key and 9{,}525 unsigned verifications, again with 0 accepts.

\subsection{Content-bound signatures}
\label{app:binding}

The default verifier authenticates the signer and the message (Section~\ref{sec:threat}). This subsection describes an optional mode that also binds the signature to the host audio, so that a payload moved into other audio no longer verifies.

\paragraph{Construction.}
The content feature $w\in\{0,1\}^{511}$ is computed from audio alone, with the same non-overlapping 2048-point STFT as the embedder. It pools the power of bins 3 to 58 (about 65 to 1{,}270\,Hz) into 16 bands and 32 time segments, takes the logarithm and a two-dimensional DCT, drops the DC term, and keeps the sign of the remaining 511 coefficients.
These bins lie below both embedding bands of the WB profile (phase bins 60 to 299, magnitude bins 100 to 339), so signing barely moves $w$. Over the 200 signed donors used below, the signed PCM16 clip differed from its host by 0 bits at the median and by at most 5 bits.

The signer uses a code-offset secure sketch with a binary BCH code of length 511 and designed distance 25, which corrects $T=12$ bit errors. It draws a pseudorandom codeword $q$ from a seed derived from the public key and the host PCM, and publishes the helper $h=w\oplus q$. The signature covers the message and the content:
\begin{align}
S &= \text{Ed25519.Sign}\big(K_\text{priv},\ \lambda\,\|\,\text{len}_{16}(M)\,\|\,M\,\|\,E\big),\\
E &= (v,f,T)\,\|\,\text{SHA-256}(w)\,\|\,h,
\end{align}
where $\lambda$ is a fixed domain tag and $(v,f,T)$ encode the format version, the feature and the radius. The extension $E$ has 99 bytes, and the RS-coded payload grows from 145 to 244 bytes for our 49-byte message.
The verifier first recovers and checks $S$ as in Section~\ref{sec:extract}. It then recomputes $w'$ from the received audio, decodes $w'\oplus h$ to a codeword $q'$ within distance $T$, and accepts the content only if $\text{SHA-256}(q'\oplus h)$ equals the signed digest. The verifier reports one of three outcomes: \emph{verified-bound}, \emph{content mismatch} (a valid signature whose content check fails), or \emph{not verified}. Only the first is an acceptance.

\paragraph{Why the signature must cover the feature.}
A variant that leaves $h$ unsigned and signs only the message and $\text{SHA-256}(q)$ does not bind content. An attacker who knows $w$ recovers $q=h\oplus w$ and publishes $h'=w_\text{target}\oplus q$ for a new target, which decodes to the same $q$. On synthetic features, this helper substitution produced 20 false accepts in 20 trials. When the signature covers $\text{SHA-256}(w)$ instead, the same substitution was rejected in 20 of 20 trials, because the reconstructed feature no longer matches the signed digest.
This is an algebraic check on random features, not an audio measurement. We report it because the weaker form is a natural first design. Our construction signs both $\text{SHA-256}(w)$ and $h$.

\paragraph{Operating point.}
We fixed the feature and the radius before any end-to-end run, from a screen of 5{,}607 ten-second clips: the 1{,}905 test-clean clips and 3{,}702 clips from other speech, music and environmental corpora.
The primary pre-registered rule asked for a radius of at least three times the worst benign feature drift, with no distinct-PCM pair within twice the radius. No feature met it. A fallback rule, registered before the screen finished, chose the feature with the smallest worst-case benign drift (61 bits, against 105 and 150 for the two alternatives). It then chose the smallest radius on the grid with a benign miss rate of at most 1\% and no unexplained distinct-PCM pair within $2T$. This gave $T=12$, with a screened miss rate of 0.3\% (117 of 39{,}249 clip-condition rows). All 277 distinct-PCM pairs within $T$ in the screen were copies of the same recording or shared a source by metadata.
Because test-clean is part of the screen, the operating point is not held out.

\paragraph{Threat model.}
The attacker holds the signed PCM16 donor, the public key, the public code and the full 5{,}607-clip corpus. The attacker does not hold the private key or the unmarked donor. We test two attacks. The naive transplant extracts the bound payload from the donor, repair it with the RS code, and re-embed it into an unrelated target with the public embedder. The nearest-host search first computes $w$ from the signed donor. For each donor it then picks the eligible corpus clip with the smallest Hamming distance to $w$, and transplants onto that clip. A clip is eligible if its PCM differs from the donor's, if it does not share a source by metadata, and if a waveform screen does not mark it as a copy of the donor's recording. The screen uses a normalized cross-correlation threshold of 0.90 that was pre-registered in the corpus study.

\paragraph{Robustness and quality cost.} On the 1{,}905 test-clean clips, the content-bound mode verifies $0.961$ under MP3 at 128\,kbps, $0.980$ under FLAC, $0.978$ after resampling to 16\,kHz and $0.975$ after an 8\,kHz low-pass, at a paired identity PESQ cost of $0.037$ [$0.033$, $0.040$] relative to the default mode. Its 45{,}720 wrong-key, unsigned and random-payload verifications produced no acceptance at the signature or the content level.

\paragraph{Naive transplant.}
We used 200 disjoint cross-speaker donor--target pairs of test-clean clips (Fig.~\ref{fig:binding}a). The attacker extracted a bound payload from 195 donors. The five failures are the same clips whose signed donor does not verify. The transplanted signature still verified on 188 of 200 targets, but the content check passed on none of them (0 of 200, 95\% upper bound 0.018). The result was the same after MP3 128k (0 of 200). The verifier reported all 188 cases as content mismatch. Donor and target features differed by 227 to 278 of 511 bits (median 252).

\paragraph{Nearest-host search.}
The search found no usable host (Fig.~\ref{fig:binding}b). After copies of the donor's recording were excluded, the nearest eligible host was 192 to 218 bits from the donor feature (median 210). No donor had an eligible host within $T$ or within $2T$. All 195 executable attacks were rejected (195 of 195, 95\% lower bound 0.981): 191 as content mismatch and 4 as not verified. The other five donors carry no publicly extractable payload, so the attacker has nothing to transplant. As a control, we also transplanted onto the 8 excluded copies. They belong to 4 donors and sit at feature distance 0. All 8 were accepted. This is the intended behavior, since these hosts carry the same recording. It also shows that the verifier rejects foreign content, not every transplant.

\paragraph{Scope.} These results cover two non-adaptive attackers on one corpus, with fixed keys, one 49-byte message and LibriSpeech donors. We draw no entropy or collision-probability claim from the feature distances.

\subsection{Correspondence to other evaluation suites}
\label{app:corr}

Table~\ref{tab:corr} maps each transformation we measured to the closest condition in four evaluation traditions: the evaluation-time augmentations of AudioSeal~\cite{sanroman2024audioseal} (training-time values in parentheses where they differ), the no-box, black-box and white-box perturbations of AudioMarkBench~\cite{adversarial_audio_survey}, the re-recording and signal distortions of DeAR~\cite{liu2023dearRR}, and the physical-access (replay) condition of ASVspoof 2019~\cite{asvspoof2019}. A dash means that the suite has no corresponding condition. Parameters are those printed in the cited papers. The last row lists conditions of these suites that our battery does not include.

\begin{table}[tbp]
\centering
\caption{Our transformations and their closest counterparts in AudioSeal, AudioMarkBench (AMB), DeAR and ASVspoof 2019 PA.}
\label{tab:corr}
\footnotesize
\setlength{\tabcolsep}{2.5pt}
\begin{tabular}{@{}>{\raggedright\arraybackslash}p{0.17\linewidth} >{\raggedright\arraybackslash}p{0.18\linewidth} >{\raggedright\arraybackslash}p{0.18\linewidth} >{\raggedright\arraybackslash}p{0.21\linewidth} >{\raggedright\arraybackslash}p{0.20\linewidth}@{}}
\toprule
Ours & AudioSeal & AMB & DeAR & ASVspoof 2019 PA \\
\midrule
MP3 32--256\,kbps & MP3 32\,kbps (train 128) & MP3 8--40\,kbps & MP3 64, 128\,kbps & -- \\
OGG-Vorbis 48--192\,kbps & -- & -- & -- & -- \\
AAC 32--192\,kbps & AAC 64\,kbps (train 128) & -- & -- & -- \\
Opus 12--128\,kbps & -- & Opus 16--256\,kbps & -- & -- \\
FLAC & -- & -- & -- & -- \\
Resampling via 8--48\,kHz & resample to 32\,kHz & -- & resample to 90\% and back & -- \\
Low-pass 4--8\,kHz & low-pass 500\,Hz (train 5\,kHz) & low-pass, cutoff ratio 0.1--0.5 & low-pass 4\,kHz & -- \\
High-pass 300\,Hz & high-pass 1.5\,kHz (train 500\,Hz) & high-pass, cutoff ratio 0.1--0.5 & high-pass 1\,kHz & -- \\
Gain $\times$0.5, $\times$2 & boost $\times$10, duck $\times$0.1 (train 1.2, 0.8) & -- & amplitude to 90\% & -- \\
Tail, head and middle cropping & -- & -- & -- & -- \\
Integer time shift & -- & -- & -- & -- \\
Time stretch, speed 0.90--1.10 & speed 1.25 (train 0.9--1.1) & time stretch 0.7--1.5 & -- & -- \\
Pitch shift & -- & -- & -- & -- \\
White noise, 50--0\,dB SNR & white noise, std 0.05 (train 0.001) & Gaussian noise 5--40\,dB & Gaussian noise 10--25\,dB & -- \\
Pink, brown noise & pink noise, std 0.1 (train 0.01) & background noise 5--40\,dB & -- & -- \\
Requantization 12--4 bit & -- & quantization, 4--64 levels & requantization 8 bit & -- \\
Packet loss 1--20\% & -- & -- & dropout of one sample in 100 & -- \\
Echo 20--100\,ms & echo 0.5\,s, volume 0.5 (train 0.1--0.5\,s, volume 0.1--0.5) & echo, delay 0.1--0.9\,s & -- & -- \\
Reverberation, measured room responses & -- & -- & conference-room impulse responses at several distances & reverberant rooms (3 categories; see replay row) \\
Classical denoisers, level processing & -- & -- & -- & -- \\
EnCodec 1.5--24\,kbps & EnCodec, $n_q{=}16$ & EnCodec 1.5--24; SoundStream & -- & -- \\
Vocoder and voice conversion & -- & -- & -- & -- \\
Record-and-replay (simulated) & -- & -- & loudspeaker to microphone or phone, 5--100\,cm & simulated replay: 27 acoustic (room size, reverberation, talker distance) and 9 replay configurations \\
Chains of 2--5 stages & -- & combined EnCodec, MP3 16\,kbps, noise 20\,dB & -- & -- \\
PGD, square search (both channels) & -- & white-box PGD (SNR 20--60\,dB); black-box square, HSJA & -- & -- \\
Header-targeted attack & -- & -- & -- & -- \\
Payload transplant & -- & forgery goal & -- & -- \\
\midrule
Not in our battery & band-pass, smooth & smooth, SoundStream, HSJA, I-FGSM & median filter & -- \\
\bottomrule
\end{tabular}
\end{table}

\section{Neural baselines, capacity and coexistence}
\label{app:baselines}

\paragraph{Baselines.} Table~\ref{tab:sota} places APC next to published and reproduced neural watermarks. We reproduced AudioSeal and WavMark at their native 16\,kHz with their released checkpoints on 200 LibriSpeech clips, and again on all 1{,}905 benchmark clips, where they reach PESQ $4.45$ and $4.12$ and recover their messages exactly on every clip, within 0.1 of the published values. Published neural payloads are 16 bits for AudioSeal, 32 bits for WavMark, and 32 or 40 bits for SilentCipher, not 64-byte signatures.

\begin{table}[tbp]
\centering
\caption{APC and neural baselines under MP3 compression. ``P'' = PESQ, ``dB'' = SDR; ``Sig.'' marks public-key signatures; ``T-free'' marks no training data and no GPU. Published rates differ: WavMark 64, AudioSeal 32, SilentCipher 128\,kbps; reproductions ($^{\dagger}$, 200 LibriSpeech clips at 16\,kHz) and APC use 128\,kbps. The APC row is the 1{,}905-clip benchmark of Table~\ref{tab:results}; on those 1{,}905 clips the reproductions reach PESQ 4.45 and 4.12.}
\label{tab:sota}
\footnotesize
\setlength{\tabcolsep}{4pt}
\begin{tabular}{@{}l c c c c@{}}
\toprule
Method & Qual. & MP3 & Sig. & T-free \\
\midrule
WavMark~\cite{chen2023wavmark}                       & 4.21\,P & $\!\approx\!$100\% & \texttimes & \texttimes \\
AudioSeal~\cite{sanroman2024audioseal}               & 4.47\,P & $\!\approx\!$100\% & \texttimes & \texttimes \\
SilentCipher~\cite{singh2024silentcipher}            & 47\,dB  & 100\%              & \texttimes & \texttimes \\
WavMark$^{\dagger}$~\cite{chen2023wavmark}           & 4.11\,P & 100.0\%            & \texttimes & \texttimes \\
AudioSeal$^{\dagger}$~\cite{sanroman2024audioseal}   & 4.44\,P & 100.0\%            & \texttimes & \texttimes \\
\midrule
APC                                                  & 3.82\,P & 99.7\%             & \checkmark & \checkmark \\
\bottomrule
\end{tabular}
\end{table}

\begin{figure}[tbp]
  \centering
  \includegraphics[width=\linewidth]{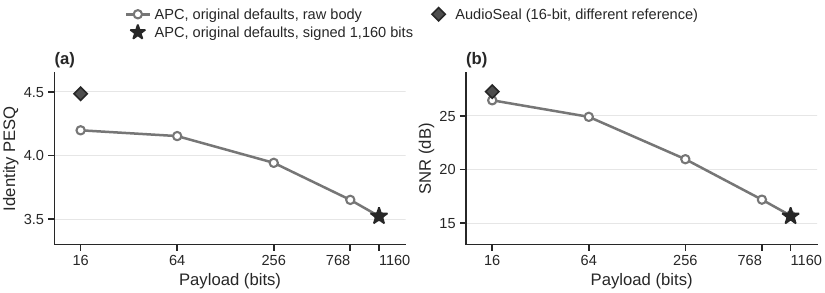}
  \caption{Identity quality against payload size with the original defaults (TED-LIUM, $n{=}30$ clips from 10 talks). (a) PESQ and (b) SNR of the marked clip relative to the original. Circles: raw payload body of the given size, without Reed--Solomon parity or signature. Star: the full signed 1{,}160-bit payload. Diamond: AudioSeal with its 16-bit message, measured against the 16\,kHz round trip of the original. At 16 bits APC reaches PESQ 4.20, close to AudioSeal's 4.49, which is measured against a different reference.}
  \label{fig:capacity}
\end{figure}

\paragraph{Capacity.} Fig.~\ref{fig:capacity} sweeps the raw APC payload from 16 to 1{,}160 bits on TED-LIUM ($n{=}30$; original defaults). At 16 bits APC reaches PESQ $4.20$ and SNR $26.4$\,dB. AudioSeal reaches $4.49$ and $27.3$\,dB, measured against the 16\,kHz round trip of the original, whereas APC is measured against the original itself, so the 0.29-PESQ gap compares slightly different references. The signed 1{,}160-bit payload sits at PESQ $3.52$ and $15.6$\,dB. The smallest self-contained signed payload (Ed25519, an empty message, RS parity) is 768 bits, $48\times$ a 16-bit neural message.

\paragraph{Chained signing and minting.} A chained layer that lets a 16-bit neural message index a signed record inherits the neural mark's exact recovery. On the same TED-LIUM clips, AudioSeal recovers its message exactly (1.000) under identity, MP3 at 128 and 64\,kbps, 16\,kHz resampling and 20\% cropping, and signed APC also verifies at $1.000$ under each of them; the chained layer, however, needs a trusted registry and a lookup for every verification, whereas APC verifies offline. Minting differs: with the released AudioSeal generator we minted valid AudioSeal marks on all 30 clips, whereas APC marks signed with an attacker key verified on none.

\paragraph{Coexistence.} We stacked APC and AudioSeal (16-bit message, detector on CPU) on the first 300 clips of the main benchmark and compared the stack with a matched control that replaces the AudioSeal step by the 16\,kHz round trip it implies. With AudioSeal applied first and APC second, the order we recommend (Section~\ref{sec:discuss}), AudioSeal recovers its 16-bit message exactly on 300 of 300 clips under the eight benchmark attacks, MP3 at 64\,kbps and 3\% packet loss, and APC's verified counts change by $0, -4, -4, 0, 0, 0, 0, 0$ relative to the control (identity, MP3, OGG, FLAC, resampling, crop 10\%, crop 20\%, low-pass), so APC shows no systematic loss.

\section{Reproducibility and compute}
\label{app:repro}

\paragraph{Release.} Code and data: \url{https://gy19a.github.io/asymmetric-phase-coding-audio-watermarking-js/}. The release contains (i) the APC reference implementation in Python and a Rust implementation whose verification results matched the Python reference on all 100 clips of the latency benchmark; (ii) the format specification (version 1) and the conformance vectors; (iii) the evaluation code and fixed keys used for every experiment; (iv) the clip list of the LibriSpeech benchmark; (v) the pre-registered design and per-clip results of every experiment; and (vi) the code, keys and results of the measurements with the original defaults (Appendices~\ref{app:quality}, \ref{app:adv} and~\ref{app:baselines}).

\paragraph{Hardware.} The measurements with the original defaults and the baseline reproductions on 200 clips were computed on a workstation with an AMD Ryzen 9 5900XT CPU and NVIDIA RTX A4000 GPUs. The other experiments ran inside an Apptainer container
on 40-core Intel Xeon Gold 6230 nodes and an Intel Xeon Platinum 8480+ node with NVIDIA H100 GPUs. The runs whose results we report used about 2{,}238 core-hours
and about 4.6 H100 GPU-hours.

\paragraph{Latency.} The 49-byte message and its signature are carried at a net 92\,bit/s (115 bytes per 10\,s; with RS parity, 145 bytes or 116\,bit/s). Table~\ref{tab:latency} gives single-thread latency on one pinned core of an Intel Xeon Platinum 8480+ with all thread pools set to one thread (10\,s clip at 44.1\,kHz; 100 clips of the main benchmark, 3 warm-up runs and 3 timed repetitions per clip, $n{=}300$ per cell). The command-line numbers include process start-up and file input and output. Verification of a marked clip has a 95th percentile of 9.52\,ms, because five of the 100 clips need the length search. The numbers describe this code on this CPU under this load; they do not include the optional shift search.

\begin{table}[tbp]
\centering
\caption{Single-thread latency (ms), median [interquartile range], Intel Xeon Platinum 8480+.}
\label{tab:latency}
\footnotesize
\IfFileExists{generated/tab_latency.tex}{\begin{tabular}{@{}lccc@{}}
\toprule
Implementation & Sign & Verify, marked & Verify, unmarked \\
\midrule
Python reference & 12.35 [12.22, 12.52] & 3.17 [3.15, 3.23] & 320.84 [318.48, 324.03] \\
Rust library & 2.62 [2.60, 2.70] & 1.51 [1.50, 1.53] & -- \\
Rust CLI, whole process & 19.44 [17.96, 21.71] & 8.01 [7.81, 8.43] & 22.28 [21.82, 22.99] \\
Rust CLI, no read-back check & 13.26 [12.84, 13.96] & -- & -- \\
\bottomrule
\end{tabular}
}{\textbf{[generated/tab\_latency.tex missing]}}
\end{table}

\section{Extended related work}
\label{app:related}

Table~\ref{tab:comparison} gives the qualitative positioning of APC among classical and neural watermarks. Table~\ref{tab:five} compares APC with five recent learned watermarks on where the mark lives, whether verification uses a public-key signature, whether the mark was evaluated after re-encoding by an external neural codec, what must be trained, and the payload. Entries follow the cited papers. None of the five uses a public-key signature, so none gives non-repudiation; only the token-domain method was evaluated after external neural-codec re-encoding.

\begin{table}[tbp]
\centering
\caption{Qualitative positioning. Rob./Imp./Sig./T-free = robustness, imperceptibility, public-key signature, training-free. Numeric comparison appears in Table~\ref{tab:sota}.}
\label{tab:comparison}
\footnotesize
\setlength{\tabcolsep}{4pt}
\begin{tabular}{@{}l c c c c@{}}
\toprule
Method & Rob. & Imp. & Sig. & T-free \\
\midrule
LSB                                        & \texttimes & \checkmark & \texttimes & \checkmark \\
Echo~\cite{gruhl1996echo}                  & $\sim$     & $\sim$     & \texttimes & \checkmark \\
Spread Spec.~\cite{cox1997}                & \checkmark & $\sim$     & \texttimes & \checkmark \\
Phase~\cite{bender1996techniques}          & $\sim$     & \checkmark & \texttimes & \checkmark \\
WavMark~\cite{chen2023wavmark}             & \checkmark & \checkmark & \texttimes & \texttimes \\
AudioSeal~\cite{sanroman2024audioseal}     & \checkmark & \checkmark & \texttimes & \texttimes \\
SilentCipher~\cite{singh2024silentcipher}  & \checkmark & \checkmark & \texttimes & \texttimes \\
\textbf{APC (ours)}                        & $\sim$     & \checkmark & \checkmark & \checkmark \\
\bottomrule
\end{tabular}
\end{table}

\begin{table}[tbp]
\centering
\caption{APC and five learned watermarks beyond the signal domain.}
\label{tab:five}
\footnotesize
\setlength{\tabcolsep}{3pt}
\begin{tabular}{@{}>{\raggedright\arraybackslash}p{0.17\linewidth} >{\raggedright\arraybackslash}p{0.18\linewidth} >{\raggedright\arraybackslash}p{0.08\linewidth} >{\raggedright\arraybackslash}p{0.23\linewidth} >{\raggedright\arraybackslash}p{0.13\linewidth} >{\raggedright\arraybackslash}p{0.09\linewidth}@{}}
\toprule
Work & Where the mark lives & Public-key signature & External neural-codec re-encoding & Training & Payload \\
\midrule
P2Mark~\cite{ren2025p2mark} & model parameters (LoRA adapter in the vocoder or codec decoder) & no & not evaluated (MP3, AAC 128\,kbps only) & adapter fine-tuning & 16 bits (32 in an ablation) \\
WMCodec~\cite{zhou2025wmcodec} & codec latent before quantization & no & not evaluated (own codec at 3--12\,kbps) & end-to-end codec training & 16 bits \\
TraceableSpeech~\cite{zhou2024traceablespeech} & codec latent of a codec-LM TTS & no & not evaluated & end-to-end training & 16 bits (up to 24 in a capacity analysis) \\
Hidden in Plain Tokens~\cite{milis2026hidden} & token sampling of an autoregressive audio model & no & evaluated (DAC, EnCodec, SpeechTokenizer, FaCodec); detection weakens but stays far above the 1\% FPR threshold & none (clustering of retokenization errors) & zero-bit detection \\
GROOT~\cite{liu2024groot} & input latent of a diffusion vocoder & no & not evaluated & watermark encoder and decoder only & 100 bits \\
\midrule
APC (ours) & STFT phase and log-magnitude of the waveform & Ed25519 & evaluated (Section~\ref{sec:limits}) & none & 1{,}160-bit signed codeword \\
\bottomrule
\end{tabular}
\end{table}

\newpage
\section*{NeurIPS Paper Checklist}

\begin{enumerate}

\item {\bf Claims}
    \item[] Question: Do the main claims made in the abstract and introduction accurately reflect the paper's contributions and scope?
    \item[] Answer: \answerYes{}
    \item[] Justification: The abstract and the contributions in Section~\ref{sec:intro} match the results: the guarantee (Section~\ref{sec:threat}), the benchmark on 1{,}905 test-clean clips and 2{,}048 corpus clips, the design ablation and the measured transport envelope (Section~\ref{sec:exp}), with the boundaries stated in Section~\ref{sec:limits}; every number is reported with its sample size and corpus.

\item {\bf Limitations}
    \item[] Question: Does the paper discuss the limitations of the work performed by the authors?
    \item[] Answer: \answerYes{}
    \item[] Justification: Section~\ref{sec:limits} states four limits. Neural codecs, speech resynthesis and time-scale changes remain limits, and the decoder-side search does not guarantee robustness to speed changes, pitch shifts, reverberation or playback and recapture. Robustness to noise, requantization, denoising and repeated transcoding depends on the strength and configuration of the processing. Adaptive payload perturbations aimed at the public layout can make verification fail with small changes in objective quality; this limits erasure resistance and is not a forgery. Lowering the embedding strength trades quality against robustness, and the platform chains were simulated.

\item {\bf Theory assumptions and proofs}
    \item[] Question: For each theoretical result, does the paper provide the full set of assumptions and a complete (and correct) proof?
    \item[] Answer: \answerNA{}
    \item[] Justification: The paper makes empirical claims and no new theoretical results. The statement that forging a valid mark requires forging an Ed25519 signature relies on the published security of Ed25519~\cite{rfc8032,bernstein2012ed25519,nist2023digital,brendel2021ed25519}; it is not a theorem of our own.

    \item {\bf Experimental result reproducibility}
    \item[] Question: Does the paper fully disclose all the information needed to reproduce the main experimental results of the paper to the extent that it affects the main claims and/or conclusions of the paper (regardless of whether the code and data are provided or not)?
    \item[] Answer: \answerYes{}
    \item[] Justification: Section~\ref{sec:method} and Appendix~\ref{app:format} give every coder parameter and the switches used by the ablation; Appendices~\ref{app:data} and~\ref{app:protocol} give the deterministic benchmark build, the attacks with their FFmpeg settings, the metrics and the statistical tests.

\item {\bf Open access to data and code}
    \item[] Question: Does the paper provide open access to the data and code, with sufficient instructions to faithfully reproduce the main experimental results, as described in supplemental material?
    \item[] Answer: \answerYes{}
    \item[] Justification: Code and data: \url{https://gy19a.github.io/asymmetric-phase-coding-audio-watermarking-js/}. Appendix~\ref{app:repro} lists what the release contains: the APC reference implementations in Python and Rust, the format specification with conformance vectors, the evaluation code, the clip lists, the fixed evaluation keys, and the pre-registered experiment designs with their results. LibriSpeech is available under CC BY 4.0; the other corpora are public, and 15 of the 18 corpus sets identify each clip by source, split and row identifier (Appendix~\ref{app:data}).

\item {\bf Experimental setting/details}
    \item[] Question: Does the paper specify all the training and test details (e.g., data splits, hyperparameters, how they were chosen, type of optimizer) necessary to understand the results?
    \item[] Answer: \answerYes{}
    \item[] Justification: APC is training-free, so there is no training split, optimizer or learning rate. All coder parameters are fixed (Section~\ref{sec:method}, Appendix~\ref{app:format}). The evaluation keys, message, attacks and clip selection are described in Section~\ref{sec:exp} and Appendices~\ref{app:data} and~\ref{app:protocol}. The AudioSeal and WavMark reproductions used their released checkpoints at their native 16\,kHz on 200 LibriSpeech clips and on the 1{,}905 benchmark clips; SilentCipher was not run, and only its published numbers are cited.

\item {\bf Experiment statistical significance}
    \item[] Question: Does the paper report error bars suitably and correctly defined or other appropriate information about the statistical significance of the experiments?
    \item[] Answer: \answerYes{}
    \item[] Justification: Verification rates are reported with 95\% Wilson or exact Clopper--Pearson intervals. Ablation differences are paired on the same clips and carry paired bootstrap 95\% intervals and exact McNemar tests; quality differences carry paired bootstrap intervals. Every number is stated with its sample size. The smaller studies with the original defaults ($n{=}30$--$150$) are labeled with their corpus, configuration and $n$.

\item {\bf Experiments compute resources}
    \item[] Question: For each experiment, does the paper provide sufficient information on the computer resources (type of compute workers, memory, time of execution) needed to reproduce the experiments?
    \item[] Answer: \answerYes{}
    \item[] Justification: APC needs no GPU; single-thread latency was measured on one pinned core of an Intel Xeon Platinum 8480+ (Section~\ref{sec:cost}). Appendix~\ref{app:repro} gives the hardware of all runs and the compute of the runs whose results we report: about 2{,}238 core-hours and about 4.6 H100 GPU-hours.

\item {\bf Code of ethics}
    \item[] Question: Does the research conducted in the paper conform, in every respect, with the NeurIPS Code of Ethics \url{https://neurips.cc/public/EthicsGuidelines}?
    \item[] Answer: \answerYes{}
    \item[] Justification: The work evaluates public speech, music and environmental-audio corpora and uses standard cryptographic primitives (Ed25519, Reed--Solomon). We collected no new recordings or annotations from people. The method is defensive: it lets a third party check who signed the message carried by a recording. We state its limits, including payload transplant and erasure, so that a verification is not read as more than it proves. We have reviewed the NeurIPS Code of Ethics and find the research consistent with it.

\item {\bf Broader impacts}
    \item[] Question: Does the paper discuss both potential positive societal impacts and negative societal impacts of the work performed?
    \item[] Answer: \answerYes{}
    \item[] Justification: \emph{Positive}: APC gives a third party an offline check of who signed the message carried by an audio file, which supports provenance pipelines against voice-cloning fraud. \emph{Negative}: a valid mark can be over-trusted. It proves the signer and the message, not that the audio is the audio that was signed, because a payload can be transplanted into other audio (Section~\ref{sec:threat}). A compromised private key lets an attacker sign arbitrary audio, and a key holder can sign synthetic or edited audio, so a valid mark establishes accountability rather than truthfulness. Erasure is always possible, and an erased clip is indistinguishable from unmarked audio. A verifier that trusts the unprotected stream header allows erasure at no measurable quality cost; APC's verifier closes this by construction. Mitigations are standard key management, pairing APC with a container-level manifest, and the optional content-bound mode of Appendix~\ref{app:binding}, at a robustness cost.

\item {\bf Safeguards}
    \item[] Question: Does the paper describe safeguards that have been put in place for responsible release of data or models that have a high risk for misuse (e.g., pre-trained language models, image generators, or scraped datasets)?
    \item[] Answer: \answerNA{}
    \item[] Justification: The paper releases a deterministic, training-free signing and verification algorithm and its evaluation code. It releases no generative model, pre-trained large model or scraped data. The released code embeds only signatures made with a private key that the user controls.

\item {\bf Licenses for existing assets}
    \item[] Question: Are the creators or original owners of assets (e.g., code, data, models), used in the paper, properly credited and are the license and terms of use explicitly mentioned and properly respected?
    \item[] Answer: \answerYes{}
    \item[] Justification: Every dataset and pretrained model is cited, and Appendix~\ref{app:data} (Tables~\ref{tab:asset-licenses} and~\ref{tab:model-licenses}) lists the version used, the source, the license stated by its creators and the software tools with their licenses. Assets with restrictions (TED-LIUM~3, MUSDB18-HQ, FMA, AudioSet, VoxConverse, Common Voice, GTZAN, the MIT IR Survey), the DNS64 model of \texttt{denoiser} (CC~BY-NC~4.0) and the \texttt{pesq}~0.0.4 package are used for non-commercial academic evaluation only, and no audio from the restricted corpora is redistributed; SilentCipher was not run.

\item {\bf New assets}
    \item[] Question: Are new assets introduced in the paper well documented and is the documentation provided alongside the assets?
    \item[] Answer: \answerYes{}
    \item[] Justification: The new assets are the APC reference implementation, the format specification with conformance vectors, the evaluation code, and the clip list of the rebuilt LibriSpeech \texttt{test-clean} benchmark. They are documented in the release (\url{https://gy19a.github.io/asymmetric-phase-coding-audio-watermarking-js/}), which lists dependencies, the exact commands behind each table and the license of each component.

\item {\bf Crowdsourcing and research with human subjects}
    \item[] Question: For crowdsourcing experiments and research with human subjects, does the paper include the full text of instructions given to participants and screenshots, if applicable, as well as details about compensation (if any)?
    \item[] Answer: \answerNA{}
    \item[] Justification: No crowdsourcing or research with human subjects was conducted, and no listening test was run. All quality statements rest on objective metrics (PESQ, STOI, SNR, LSD) and an automatic speaker verifier.

\item {\bf Institutional review board (IRB) approvals or equivalent for research with human subjects}
    \item[] Question: Does the paper describe potential risks incurred by study participants, whether such risks were disclosed to the subjects, and whether Institutional Review Board (IRB) approvals (or an equivalent approval/review based on the requirements of your country or institution) were obtained?
    \item[] Answer: \answerNA{}
    \item[] Justification: No research with human subjects was conducted and no new data were collected from people. All recordings come from public corpora released by their creators.

\item {\bf Declaration of LLM usage}
    \item[] Question: Does the paper describe the usage of LLMs if it is an important, original, or non-standard component of the core methods in this research? Note that if the LLM is used only for writing, editing, or formatting purposes and does \emph{not} impact the core methodology, scientific rigor, or originality of the research, declaration is not required.
    \item[] Answer: \answerNA{}
    \item[] Justification: The core method (asymmetric phase coding + magnitude-QIM + Ed25519 + Reed--Solomon + soft decoding) does not use any large language model as a methodological component. LLMs were not used to design, implement, train, or evaluate the watermark, nor in the benchmark or attack pipelines.

\end{enumerate}

\end{document}